\documentclass[fleqn,usenatbib]{mnras}

\usepackage{newtxtext,newtxmath}

\usepackage[T1]{fontenc}

\DeclareRobustCommand{\VAN}[3]{#2}
\let\VANthebibliography\thebibliography
\def\thebibliography{\DeclareRobustCommand{\VAN}[3]{##3}\VANthebibliography}

\usepackage{graphicx}	
\usepackage{amsmath}	
\usepackage[flushleft]{threeparttable}
\usepackage{siunitx}
\usepackage{booktabs, caption}
\usepackage{float}
\usepackage{pdflscape}
\usepackage{ulem}
\usepackage{xcolor}
\usepackage{tcolorbox}

\title[ROA Models of Quasar Lightcurves]{Bayesian Analysis of Quasar Lightcurves with a Running Optimal Average: New Time Delay Measurements of COSMOGRAIL Gravitationally Lensed Quasars}

\author[F.~R.~Donnan et al.]{
Fergus~R.~Donnan,$^{1}$\thanks{E-mail: frd3@st-andrews.ac.uk}
Keith~Horne,$^{1}$
and Juan~V.~Hern\'andez~Santisteban$^{1}$
\\
$^{1}$SUPA School of Physics and Astronomy,
University of St~Andrews,
North Haugh, 
St~Andrews KY16~9SS, Scotland, UK
}

\date{Accepted XXX. Received YYY; in original form ZZZ}

\pubyear{2021}

\begin{document}
\label{firstpage}
\pagerange{\pageref{firstpage}--\pageref{lastpage}}
\maketitle

\begin{abstract}
We present a new method of modelling time-series data based on the running optimal average (ROA). By identifying the effective number of parameters for the ROA model, in terms of the shape and width of its window function and the times and accuracies of the data, we enable a Bayesian analysis, optimising the ROA width, along with other model parameters,
by minimising the Bayesian Information Criterion (BIC)
and sampling joint posterior parameter distributions using MCMC methods.
For analysis of quasar lightcurves, our implementation of ROA modelling can
measure time delays among lightcurves at different wavelengths or from different images of a lensed quasar and, in future work, be used to inter-calibrate lightcurve data from different telescopes and estimate the shape and thus the power-density spectrum of the lightcurve.
Our noise model implements a robust treatment of outliers
and error-bar adjustments to account for additional variance
or poorly-quantified uncertainties.
Tests with simulated data
validate the parameter uncertainty estimates.
We compare ROA delay measurements with results from
cross-correlation and from JAVELIN, which models
lightcurves with a prior on the power-density spectrum.
We analyse published COSMOGRAIL lightcurves 
of multi-lensed quasar lightcurves and 
present the resulting measurements of the inter-image
time delays and detection of microlensing effects.

\end{abstract}




\section{Introduction}
Active galactic nuclei (AGN) are powered by accretion onto a super-massive black hole (SMBH), producing the brightest persistent objects in the Universe \citep[e.g.][]{Salpeter1964, LyndenBell1969, Sanders1989}. AGN are known to play a key role across cosmic time in the evolution of galaxies \citep[e.g.][]{Fabian2012, Heckman2014}, and so understanding the central regions that power AGN is crucial. Despite their extreme brightness, resolving structure close to the black hole remains a challenge. Spatially resolving these regions typically requires sub-microacrsecond resolution, a feat outwith the spatial resolution of current telescopes. Recent progress has been made relying on interferometry with instruments such as GRAVITY in the near-IR \citep{Gravity2018} resolving the the broad line region (BLR) on sub-parsec scales, or the Event Horizon Telescope \citep{EHT2019} resolving the shadow of the SMBH at the centre of M87.

AGN are known to be variable \citep[e.g. ][]{Kawaguchi1998, Dexter2011}, and while its physical origin is not fully understood, this variability can be exploited to probe the inner regions of AGN. Echo mapping or reverberation mapping \citep[e.g.][]{Blandford1982, Peterson1993, Cackett2021}, is a technique that exploits this variability as well as the finite travel time of light to dissect the accretion flow, providing a probe of the structure on a scale equivalent to sub-microarcsecond resolution. This technique is based upon variable X-ray/EUV emission originating close to the BH, that propagates outward at the speed of light and is reprocessed, either as thermal continuum emission from the disc or as broad emission lines from the BLR. 

The main aim of reverberation mapping experiments is to measure the time delay between variations in the driving and reprocessed lightcurves. In the case of thermal reprocessing in the accretion disc, the wavelength is set by the local temperature, therefore measuring the time delays as a function of wavelength, provides a test of the temperature structure of the disc as well as the size of the disc \citep[][]{Cackett2007}. Measuring the delay of the response of broad emission lines, provides an estimate of the size of the BLR and thus the mass of the BH. \citep[e.g.][]{Peterson2004}. Measuring the response of these broad lines as a function of velocity, allows the structure and kinematics of the BLR to be investigated \citep[][]{Horne2004}. For these experiments, a robust method of measuring these time delays is required.

 Time delays can also be measured between images of gravitationally lensed quasars by similarly exploiting the intrinsic variability. These can provide a direct probe of Hubble's constant, $H_0$ \citep[e.g.][]{Tewes2012}, measure the size of the BLR of the source quasar \citep[e.g.][]{Sluse2014} and enable reverberation mapping at high redshift through reconstructing rest frame lightcurves \citep[e.g. ][]{Williams2021}. Measuring these time delays also requires accounting for microlensing effects which vary the brightness of the lensed images on a longer timescale than the intrinsic quasar variability \citep[e.g. ][]{Tewes2013}. 

Typically time delays between two lightcurves have been measured using the interpolation cross-correlation function (ICCF) \citep{ICCF}. The most notable problem with this method is that it is based upon interpolating one lightcurve and treating this as the driver to measure the delay to another. This is especially a problem for unevenly sampled data or large gaps where observations may have been halted for a period of time, often resulting in poor constraints on these time delays. An alternative approach would be to model the lightcurves based on some assumptions. The most popular of these, JAVELIN \citep{JAVELIN}, models the variability with a damped random walk, and generally provides tighter uncertainties \citep[e.g.][]{Yu2020} on the measured delays as well as provides a model of the driving lightcurve. A similar approach is taken by CREAM \citep{CREAM}, which models the lightcurves based on the ``lamp-post'' model, inferring the driving lightcurve. 

In this paper we investigate using a running optimal average (ROA) to model AGN lightcurves and to then measure the time delays between lightcurves. We present the code {\sc PyROA} \footnote{\url{https://github.com/FergusDonnan/PyROA}}, which draws information from all available lightcurve data in forming the ROA model of the AGN variations.
The ROA is then normalised, shifted and scaled to fit the individual lightcurves, thus measuring the mean and rms of the variations in each lightcurve, and the time delays between them. Markov Chain Monte Carlo (MCMC) samples provide parameter estimates and uncertainties. In Section~\ref{sec:Modelling} we outline the running optimal average and the modelling process of {\sc PyROA} to measure inter-lightcurve delays. In Section~\ref{sec:Mock} we test the method with simulated data. In Section~\ref{sec:GravLensing}, we use our method to measure inter-image delays of gravitationally lensed quasars of public data from the COSMOGRAIL project \citep[][and references therein]{COSMOGRAIL}.
We conclude in Section~\ref{sec:Conclusions} with a brief summary of our findings.

\section{Lightcurve Modelling}
\label{sec:Modelling}
This section outlines a Bayesian analysis using a Running Optimal Average (ROA) to model lightcurve data. The ROA model
represents the lightcurve as a running optimal average of the data. The Bayesian Information Criterion (BIC) is then used to tune the degree of smoothing warranted by the data.
The model parameters are sampled with Markov  chain Monte Carlo (MCMC) to estimate their values and uncertainties.
Unlike methods such as cross-correlation that compare one lightcurve with another, the ROA model fits multiple lightcurve data sets simultaneously to collect all the available information in determining the lightcurve shape.

\subsection{Running Optimal Average}

The ROA model provides a smooth differentiable function
that describes the shape of a lightcurve, or other time series, along with an error envelope to quantify its uncertainty. It is defined as follows. Consider time-series data consisting of $N$ data points $D_i$, with error bars $\sigma_i$, at times $t_i$. The ROA model lightcurve $X(t)$ is an optimal (inverse-variance weighted) average of all the data, evaluated at time $t$ with a window function that diminishes the influence of data at times $t_i$ that are far from the time $t$. This is given by
\begin{equation}
\label{eqn:ROA}
    X(t) = \frac{\sum_{i=1}^{N} D_i w_i(t)}{\sum_{i=1}^{N} w_i(t)}
    \ , \quad w_i(t) = \frac{1}{\sigma_i^2} \exp\left[-\frac{1}{2} \left( \frac{t - t_i}{\Delta}\right) ^2 \right] \ ,
\end{equation}
where the weights $w_i(t)$ are described here by a Gaussian window function with width $\Delta$. Given the uncertainties $\sigma_i$ on the data $D_i$, the resulting statistical variance of the running optimal average is 
\begin{equation}
\label{eqn:ROAerrs}
    \sigma^2 \left[X(t)\right] 
    = \frac{1}{\sum_{i=1}^{N} w_i(t)}.
\end{equation}
This defines the error envelope for the ROA lightcurve model.
When data points are densely sampling compared with 
the window width $\Delta$, they are averaged with optimal inverse-variance weights. The ROA interpolates across data gaps, and extrapolates beyond the ends of the data, with an error envelope that expands appropriately, depending on the adopted shape of the window function.

Unlike other attempts to model quasar variability, the ROA does not make any assumptions about the shape of the driving lightcurve. For example JAVELIN uses a damped random walk which is then smoothed by a uniform transfer function to model the lightcurves, whereas the ROA simply calculates the shape from the data, which is already smoothed. This provides a unique insight into quasar variability compared to previous methods. 

\subsubsection{ Window Function Shape}

The ROA can be defined in terms of a generic window function shape $W(x)$, where
$x\equiv(t-t_i)/\Delta$, and we require $W(0)=1$ and $W(\pm\infty)=0$.
Perhaps the simplest option is the`top-hat' window function,
\begin{equation}
    W(x) \equiv \left\{ 
    \begin{array}{cc}
    1\ , & |x| \ < \ 1 \ ,
\\  0\ , & |x| \ > \ 1 \ .
    \end{array}
    \right. \ 
\end{equation}
This admits data within a time interval $\pm\Delta$.
Although widely used,
the resulting ROA $X(t)$ jumps whenever data enter or leave the window, and the uncertainty $\sigma\left[(X(t)\right]$ is infinite
whenever a data gap exceeds the window width $2\,\Delta$.
These undesirable features are avoided with smoothly-declining window functions such as
a Gaussian window function,
\begin{equation}
\label{eqn:Gauss}
    W(x) \equiv \exp{\left( - \frac{ x^2 }{2 \, \Delta }\right)} \ .
\end{equation}
Other shapes with wider wings affect the rate at which
data more distant in time lose influence on the ROA. For example an inverse-cosh window function
\begin{equation}
\label{eqn:1/cosh}
    W(x) \equiv \frac{ 1 }{\cosh{(x)}} \ ,
\end{equation}
has exponentially decaying wings,
and a Lorentzian window function
\begin{equation}
\label{eqn:Lorz}
    W(x) \equiv \frac{ 1 }{ 1+x^2 } \ ,
\end{equation}
has power-law $x^{-2}$ wings.
Other choices are clearly possible.

\subsubsection{ Effective number of Parameters }

The ``flexibility'' of the ROA is controlled by the window function width $\Delta$. If $\Delta$ is small, then $X(t)$ is flexible enough to follow relatively rapid variations in the data. If $\Delta$ is large, then $X(t)$ is stiffer and can follow only slower variations. In the limit $\Delta \rightarrow \infty$, the ROA model becomes a rigid constant, the optimal average of all the data. From this it is clear that the value of $\Delta$ controls the effective number of parameters of the model. A small $\Delta$ highly  flexible ROA model has many parameters.
As $\Delta \rightarrow \infty$ the number of parameters reduces to just 1.
As $\Delta \rightarrow 0$, $X(t)$ can fit the data perfectly, and thus in this limit the number of parameters becomes $N$, the number of data points. 

 Optimising $\Delta$ is important because an overly-stiff model fails to fit the data while an overly-flexible model over-fits the noisy data. 
 The balance between over-fitting and under-fitting can be achieved by a trade-off between the quality of the fit, as measured for example by $\chi^2$, and an Occam bias favouring simpler models with relatively few parameters. To implement this, we need to quantify the effective number of parameters for a given value of $\Delta$. 
 
 The number of parameters $P_i$ used by the ROA model $X(t)$ to fit a single datum $D_i\pm \sigma_i$ at time $t_i$ is given by:
\begin{equation}
    P_i (\Delta) = \frac{\partial X(t_i)}{\partial D_i} = \frac{w_i(t)}{\sum_k w_k(t_i)} = \frac{1/{\sigma_i^2}}{\sum_k \frac{1}{\sigma_k^2} \exp\left[-\frac{1}{2} \left( \frac{t_i - t_k}{\Delta}\right) ^2 \right]}\ ,
\end{equation}
written here for a Gaussian window function.
Note that $P_i=1$ for an isolated data point with no other data close enough to affect the ROA. In a more densely-sampled region, $P_i$ decreases to the data point's share of the inverse-variance weights of all data close enough in time to contribute to the ROA.
For the full ROA model, $X(t)$,
the total number of parameters  sums the previous equation for $P_i$ over all $N$ data points:
\begin{equation}
\label{eqn:N_params}
    P_X (\Delta) = \sum_{i=1}^N P_i = \sum_{i=1}^N \frac{1/{\sigma_i^2}}{\sum_k \frac{1}{\sigma_k^2} \exp\left[-\frac{1}{2} \left( \frac{t_i - t_k}{\Delta}\right) ^2 \right]}\ .
\end{equation}
Knowing the number of parameters allows for the optimal value of $\Delta$ to be determined using the Bayesian Information Criterion (BIC). This is a ``badness of fit'' statistic which includes a penalty for models with too many parameters. We have released the code for simply calculating the running optimal average separately to the PyROA code for modelling the lightcurves. This code, {\sc ROA}\footnote{\url{https://github.com/FergusDonnan/Running-Optimal-Average}}, calculates the running optimal average and effective number of parameters for some given data.

The ROA model, $X(t)$, can be normalised such that  $\left< X\right>_t=0, \quad \left< X^2 \right>_t=1$. This then provides a dimensionless driving lightcurve, that can be scaled and shifted to fit to data. This is the basis of this method to model AGN lightcurves.

\subsection{Simple Model}
\label{sec:SimpleModel}
The simplest model for these lightcurves is one where the driving lightcurve is scaled, shifted, and translated in time to reproduce the flux of a given lightcurve, $i$. The following is an equation for the model flux as a function of time, $f_i(t)$, given by

\begin{equation}
\label{eqn:simple_model}
    f_{i}(t) = A_{i}X\left(t - \tau_{i}\right) + B_{i} \ ,
\end{equation}
where $A_{i}$ represents the rms flux, $B_{i}$ represents the mean flux, $\tau_{i} $ represents the time delay, and $X(t)$ is the driving lightcurve. The time delay for the first lightcurve is fixed at zero ($\tau_{1} = 0$ ) which means that the time delays are measured relative to this lightcurve.

In this simple model the assumption is that the shape of the variability given by $X(t)$, is the same for each lightcurve, calculated using all of the lightcurves shifted and stacked appropriately where the ROA is calculated from equation (\ref{eqn:ROA}). This provides the maximum information for determining the shape of the driving lightcurve. In this model, we can also add an extra variance term to the noise model as a free parameter for each lightcurve, to account for additional uncertainty not included in the original error bars. 

We fit this model using a Bayesian approach where the posterior probability of the model, $M$, given the data, $D$, ($\textrm{Pr}(M|D)$) is maximised by the best fit parameters. The natural log of this probability is given by the sum of the log prior ($\textrm{Pr}(M)$) and the log-likelihood ($\textrm{Pr}(D|M)$):
\begin{equation}
    \ln{\textrm{Pr}(M|D)}= \ln{\textrm{Pr}(M)} + \ln{\textrm{Pr}(D|M)} + \textrm{const} \ .
\end{equation}

As mentioned previously, fitting this model required the use of a statistic that includes a penalty for a running optimal average that is too flexible. As the BIC is a ``badness of fit'' statistic, it is minimised by the best fit parameters. The BIC is given by
\begin{equation}
\label{eqn:BIC}
\textrm{BIC} = -2\ln{\textrm{Pr}(D|M)} + P\ln{N} \ ,
\end{equation}
where the second term is the penalty, added to negative twice the log likelihood, which depends on the number of parameters, $P$, and the total number of data points, $N$. For our model the first term is

\begin{equation}
\label{eqn:-2lnl}
    -2\ln{\textrm{Pr}(D|M)} = \sum_{i=1}^{N_l} \left[ \sum_{j=1}^{N_i} \left[ \frac{\left(D_{j, i} - f_i\right)^2}{\sigma_{j, i}^2 + s_{i}^2 }  +   \ln\left(  \frac{\sigma_{j, i}^2 + s_{i}^2}{\sigma_{j, i}^2}\right) \right] \right] \ ,
\end{equation}

where for $N_l$ lightcurves, indexed $i = 1, 2, 3, ...$, each contain $N_i$ data points, $D_{j, i}$, with errors, $\sigma_{j, i}$. The extra error parameters, $s_i$, are added in quadrature to the original error bars. The right hand term of this equation provides a penalty for adding additional variance to the flux error measurements, which is measured relative to the original error bars to give zero when no extra variance is added.

The penalty term for the number of parameters is given by
\begin{equation}
    P\ln{N} = \sum_{i=1}^{N_l} 4\ln(N_{i}) + P_X\ln\left( \sum_{i=1}^{N_l} N_{i} \right) \ .
\end{equation}

The first term features a factor 4, for the 4 parameters per lightcurve: $A_i, B_i, \tau_i, s_i$. The number of parameters in the final term, $P_X$, depends on the free parameter $\Delta$, according to equation (\ref{eqn:N_params}). This ensures that the best fit running optimal average is the one that fits the data well with the fewest effective number of parameters.

\subsection{Fitting Procedure}
\label{sec:Fitting}
To determine the best fit parameters for this model we use the Markov chain Monte Carlo (MCMC) package, {\sc emcee} \citep{emcee}. This process samples the posterior probability of the model given the data, including prior probabilities for the parameters. The priors are chosen as uniform distributions between sensible limits e.g. $A_i$ is known to be positive. For each sample in the MCMC, the following steps are taken to calculate the posterior probability:
\begin{enumerate}
  \item First, each lightcurve is shifted by the appropriate parameters. This means each data point, $D_{j, i}$, is altered such that $D_{j, i} \rightarrow (D_{j, i} - B_i)/A_i$ and is shifted back in time by $\tau_{i}$. This has the effect of ``stacking'' the lightcurves (if the parameters are close to optimal), allowing for the running optimal average to be determined. The first lightcurve is not shifted in time ensuring that the ``stacking'' occurs on lightcurve 1. This means the time delays are measured relative to this lightcurve which removes a degeneracy in the time delays. Without this there is no reference to measure a lag from and therefore the $\tau_{i}$ parameters would be degenerate.
 \item The extra error parameters, $s_i$, are also added in quadrature to the error bars of each lightcurve, i.e $\sigma_{j, i} \rightarrow \sqrt{\sigma_{j, i}^2 + s_i^2}$.
  \item The shifted lightcurves are then merged into a single lightcurve, where the running optimal average, $X(t)$, is calculated on a fine grid of times using equation~(\ref{eqn:ROA}). The grid consists of 1000 equally spaced points, ranging over the initial and final times of the merged lightcurve. The effective number of parameters given the value of $\Delta$, is also calculated using equation (\ref{eqn:N_params}).
  \item The running optimal average, $X(t)$, is then normalised to ensure that $\left< X\right>_t=0, \, \left< X^2\right>_t=1$. This is done by subtracting the mean of $X(t)$, and then dividing by the standard deviation of $X(t)$.
  \item The model is then calculated for each lightcurve, using equation~(\ref{eqn:simple_model}). 
  \item Finally, the BIC is calculated using equation (\ref{eqn:BIC}). The negative of the BIC provides twice the log likelihood plus a penalty, therefore the relative log posterior probability is calculated by adding the twice the log prior to the negative of the BIC. i.e $2\ln{\textrm{Pr}(M)} - \textrm{BIC}$. This is the statistic calculated for each sample of the Markov chain, and maximised by the best fit solution.

\end{enumerate}

This process is repeated for a large number of samples, discarding a fraction of samples as ``burn-in''. The best fit parameters are chosen by taking the median of the posterior distributions, with uncertainties in the range from the 16\textsuperscript{th} to the 84\textsuperscript{th} percentile. The starting positions of the MCMC walkers are chosen as Gaussian random numbers centered on a chosen value. For the parameter $A_i$, the walkers are started around the rms of the individual lightcurves. For $B_i$, they are started around the mean of the individual lightcurves and for $\Delta$ they are started around 1 day - although this value can be altered depending on the data. The other parameters were started around zero. The initial time delays and $\Delta$ can be chosen depending on the data being modelled. For example, if the time delay is large and can be estimated visually (or by some other method), initialising the walkers by this estimation will reduce the ``burn-in'' required and/or prevent the Markov chain getting stuck in a local minima.

\begin{figure}
\hspace*{-0.8cm}                                                           
    \centering
	\includegraphics[width=\columnwidth]{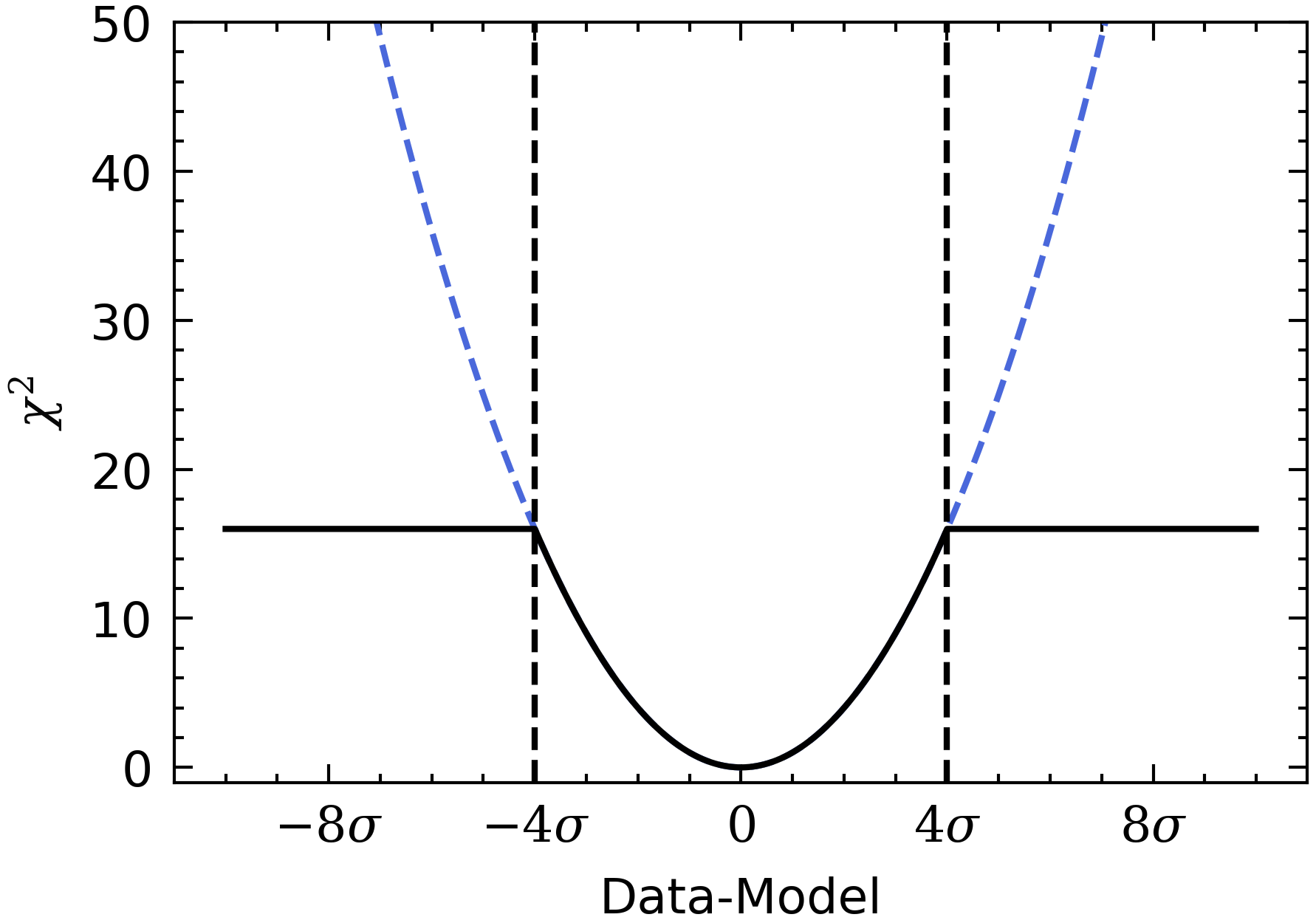}
    \caption{Demonstration of sigma clipping at a threshold of $4\sigma$. The value of $\chi^2$ is constant beyond the threshold, shown by the solid line. The dashed blue line shows $\chi^2$ without any sigma clipping.}
    \label{fig:chi^2}
\end{figure}

\subsection{Managing Outliers}
\label{sec:Outliers}
A major problem with astronomical data are outliers, caused by external processes such as cosmic rays. The optimal process of managing them is often unclear as there are many approaches. A common method is sigma clipping, where data points outwith a certain threshold ($N\sigma$) of the model, are excluded. For example, if a datum were outside a threshold of $3\sigma$, it lies outwith a probability of 99.7\%, assuming Gaussian error bars. 

Sigma clipping has the effect of creating a discontinuity in $\chi^2$ at the threshold, as $\chi^2$ drops to zero beyond the threshold. This discontinuity is undesirable as a data point slightly outwith the threshold is treated vastly differently to one slightly within the threshold. The most simple way to resolve this issue is to instead have the $\chi^2$ to become constant at the value of the threshold, demonstrated in Fig.~\ref{fig:chi^2}. This is equivalent to expanding the error bars of the outliers such that they are exactly $N\sigma$ away from the model. This method of sigma clipping was implemented into the fitting process when calculating the BIC, where we treat the $\chi^2$ term of equation (\ref{eqn:-2lnl}) as a piecewise function that is constant for data beyond the threshold. The second term is handled by expanding the error bars of points outwith the threshold to meet $N\sigma$ i.e $\sigma_{j, i}^2 + s_{i}^2$ is set to the variance required such that the datum is exactly $N\sigma$ from the model.

Other methods of sigma clipping are also possible. While $\chi^2$ is no longer discontinuous as the threshold, the first derivative is still discontinuous. A way to solve this issue would be for $\chi^2$ to become linear with the gradient set by the first derivative evaluated at the threshold. This is effectively converting $\chi^2$ into a function similar to the Median Absolute Deviation (MAD) beyond the threshold, which is much more resilient to outliers than $\chi^2$. Implementing such a solution would be relatively straightforward, if required in future applications of the model.


\section{Testing with Mock Data}
\label{sec:Mock}
\begin{figure*}
\hspace*{-0.2cm}                                                           
    \centering
	\includegraphics[width=18cm]{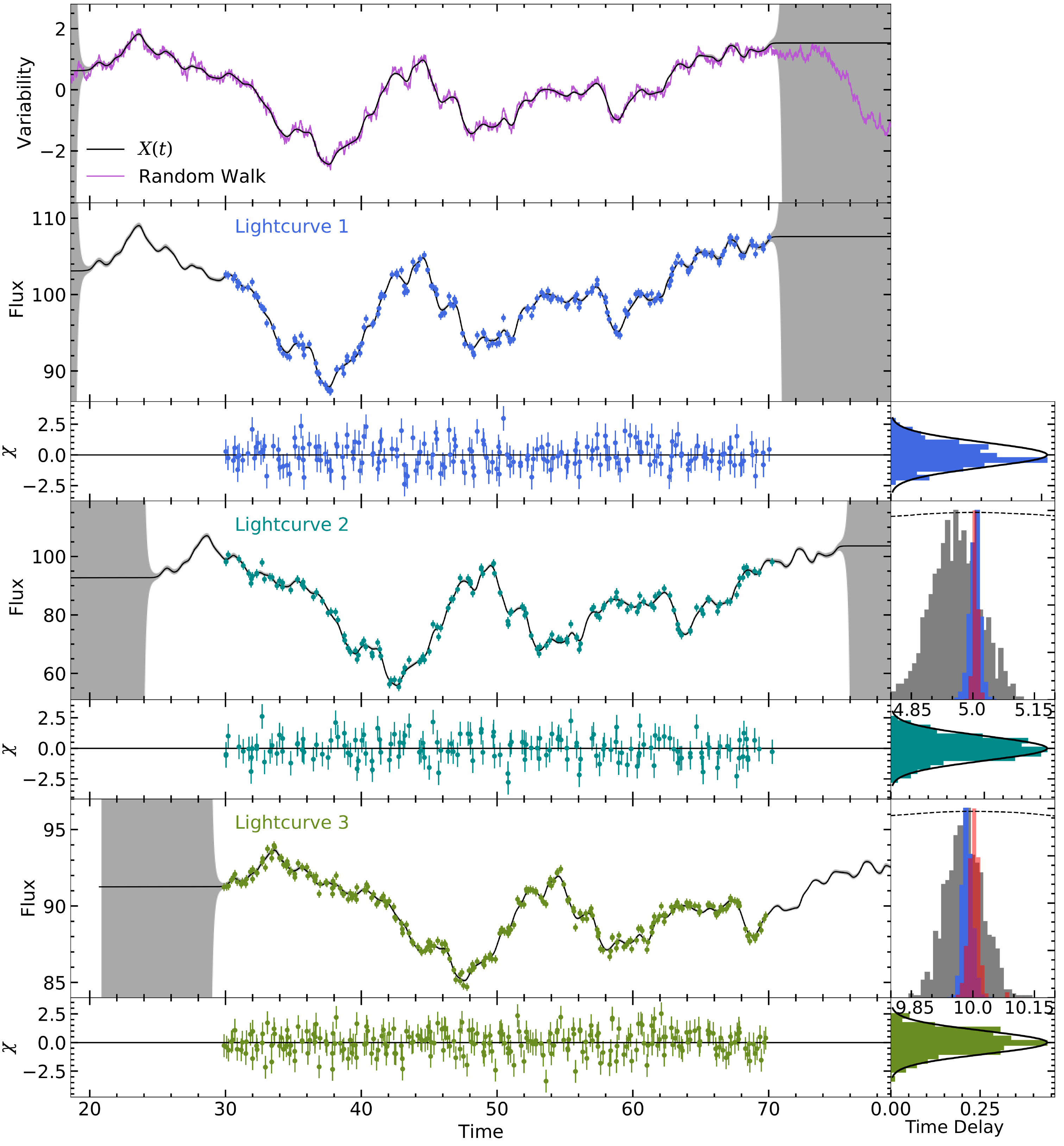}
    \caption{ Model fit to mock data, Case A. The top panel shows the random walk used to generate the data (purple) overlayed with the normalised driving lightcurve found by the model (black). The grey shaded region shows the error envelope for the running optimal average calculated using equation (\ref{eqn:ROAerrs}). The three mock lightcurves are plotted in the following panels, overlayed with the best fit model in black. The normalised residuals, $\chi$, for each lightcurve are also shown, with the color corresponding to the appropriate lightcurve. The right panels of each lightcurve show histograms of the probability distributions for the time delay. The cross-correlation centroid distribution is shown in gray, our method (PyROA) is shown in blue and JAVELIN is shown in red. The dashed line shows the cross-correlation function. The right panels of the residuals show a histogram of those normalised residuals, in comparison with the expected Gaussian distribution in black.}
    \label{fig:MockData}
\end{figure*}

To ensure the ROA model and fitting procedure is robust, we tested our algorithm with mock data where the variability is generated from a random walk. A damped random walk has been shown to describe quasar variability well \citep[e.g.][]{Kozlowski2010, MacLeod2010}, where the variability is given by a random walk on short timescales but ``damped'' on long timescales to push deviations towards the mean, with typical damping timescales on the order $\sim 200$ days \citep{MacLeod2010}. For our simulation we use a dimensionless duration of 50 and delays of 5 and 10 which if measured in days, are plausible delays for reverberation mapping studies of BLR emission line lags \citep[e.g. ][]{Grier2017} or accretion disc lags \citep[e.g. ][]{Homayouni2021}. Therefore it is suitable to use a random walk for 50 days as this is significantly shorter than the typical damping timescale of $\sim 200$ days. We first generated three mock lightcurves based on the same random walk where each is shifted in time by some true parameter. This allows us to test our method's ability to reproduce the true value of the parameters. The mock data was generated by the following.
\begin{enumerate}
    \item A random walk lightcurve was generated where each step is a Gaussian random number with a mean of 0 and a standard deviation of 1. This was done for 10000 steps over a range of times from 0 to 100.  The random walk was then normalised such that its mean is zero and rms is one. This mimics the random variability of an AGN and an example is plotted in the top panel of Fig.~\ref{fig:MockData} in purple.
    \item A set of discrete times was generated between 30 and 70 with equal spacing, consisting of 200 points for the first lightcurves, 150 for the second and 250 points for the third. A Gaussian random number with a mean of 0 and a standard deviation of 1 was then multiplied by the spacing between the times and added to the original times. This makes the spacing between the data points uneven.
    \item The normalised random walk was then scaled and shifted in time by the true parameters (shown in Table~\ref{tab:Testing}), and calculated at the times generated in the previous step.
    \item To simulate error bars, errors were chosen as some arbitrary value plus a small uniform random number to vary the sizes of the error bars across the lightcurve. 
    \item To scatter the flux values based on the error bars, they were calculated as Gaussian random numbers with a mean calculated in the 2\textsuperscript{nd} step and a standard deviation given by the error bars.
\end{enumerate}
Our benchmark model consists of three lightcurves, with the true parameters are given in Table~\ref{tab:Testing} \, labelled Case A. This was chosen to give a signal-to-noise ratio of $\sim 20$. For the first lightcurve we generated errors of 0.2, for the second we used 0.5 and 0.2 for the third. We then add a uniform random number between 0 and 0.01 for lightcurves 1 and 3 whereas for lightcurve 2 we use a uniform random number between 0 and 0.05, which varies the errors slightly. This represents a very high signal to noise case. The high signal-to-noise here is measured as how variable the source is relative to the noise of the flux measurements. This can be estimated by the ratio of the rms of the lightcurve to the mean error of the fluxes, which for this case gives a signal to noise ratio of $\sim$ 20. 

The high cadence and signal to noise are similar to typical intensive disc-reverberation mapping (IDRM) campaigns \citep[e.g.][]{Fausnaugh2016, Edelson2019, HernandezSantisteban2020, Kara2021} with facilities such as the Neil Gehrels
Swift Observatory \citep{Gehrels2004} or the Las Cumbres Observatory global telescope network \citep{Brown2013}.

These lightcurves are shown in Fig.~\ref{fig:MockData}, labelled lightcurves 1, 2, 3. The model was fitted to the data through the process described in Sec. \ref{sec:Fitting}, with 15 000 samples, 26 walkers and a burn-in of 10 000. The priors were uniform distributions between two limits given in Table~\ref{tab:Testing}.


\begin{table}
\centering
  \caption{Mock Data Results}
  \label{tab:Testing}
    \def\arraystretch{1.15}
    \setlength{\tabcolsep}{3pt}
    \begin{threeparttable}
    
\begin{tabular}{cccccc}

    \hline

    Parameter & Truth  & Prior  & Best Fit & ICCF & JAVELIN\\
    \hline
  \multicolumn{6}{c}{Case A: High Signal to Noise (Fig. \ref{fig:MockData})} \\
\cmidrule(r){1-6}

    $A_1$ & 5.0 & [0, 20] & $4.954^{+0.035}_{-0.035}$ & - & -\\
    $A_2$ & 12.0 &[0, 20] & $12.04^{+0.10}_{-0.10}$ & -& -\\
    $A_3$ & 2.0 & [0, 20]& $2.003^{+0.014}_{-0.014}$ & -& -\\
    $B_1$ & 100.0 &[0, 1000] & $100.000^{+0.033}_{-0.035}$ & -& -\\
    $B_2$ & 85.0 & [0, 1000]& $85.186^{+0.093}_{-0.096}$ & -& -\\
    $B_3$ & 90.0 &[0, 1000] & $90.003^{+0.014}_{-0.014}$ & -& -\\
    $s_1$ & 0.0& [0, 10] & $0.509^{+0.027}_{-0.026}$ & -& -\\
    $s_2$ & 0.0 & [0, 10]& $1.322^{+0.078}_{-0.069}$ & -& -\\
    $s_3$ & 0.0 & [0, 10]& $0.197^{+0.013}_{-0.013}$ & -& -\\
    $\tau_2$ & 5.0 & [-50, 50]& $5.006^{+0.011}_{-0.013}$ & $4.958^{+0.058}_{-0.062}$ & $5.0061^{+0.0059}_{-0.0052}$\\
    $\tau_3$ & 10.0 &[-50, 50] & $9.986^{+0.013}_{-0.010}$ & $9.991^{+0.048}_{-0.040}$& $10.002^{+0.012}_{-0.011}$\\
    $\Delta$ & - & [0.1, 10] &  $0.239^{+0.010}_{-0.010}$ & -& -\\
    \hline
  \multicolumn{6}{c}{Case B: Seasonal Gaps (Fig. \ref{fig:MockData2})} \\
\cmidrule(r){1-6}

    $A_1$ & 5.0 & [0, 20] & $4.953^{+0.043}_{-0.043}$ & - & -\\
    $A_2$ & 12.0 &[0, 20] & $12.57^{+0.18}_{-0.17}$ & -& -\\
    $A_3$ & 2.0 & [0, 20]& $1.990^{+0.018}_{-0.018}$ & -& -\\
    $B_1$ & 100.0 &[0, 1000] & $99.915^{+0.047}_{-0.048}$ & -& -\\
    $B_2$ & 85.0 & [0, 1000]& $84.84^{+0.11}_{-0.11}$ & -& -\\
    $B_3$ & 90.0 &[0, 1000] & $90.075^{+0.020}_{-0.020}$ & -& -\\
    $s_1$ & 0.0 & [0, 10] & $0.396^{+0.031}_{-0.030}$ & -& -\\
    $s_2$ & 0.0 & [0, 10]& $0.956^{+0.100}_{-0.093}$ & -& -\\
    $s_3$ & 0.0 & [0, 10]& $0.172^{+0.020}_{-0.020}$ & -& -\\
    $\tau_2$ & 5.0 & [-50, 50]& $5.009^{+0.018}_{-0.013}$ & $4.20^{+0.17}_{-0.15}$ & $5.016^{+0.017}_{-0.019}$\\
    $\tau_3$ & 10.0 &[-50, 50] & $9.993^{+0.021}_{-0.024}$ & $8.56^{+0.18}_{-0.20}$& $10.037^{+0.049}_{-0.048}$\\
    $\Delta$ & - & [0.1, 10] &  $0.206^{+0.013}_{-0.012}$ & -& -\\  
    \hline
  \multicolumn{6}{c}{Case C: Low Signal to Noise (Fig. \ref{fig:MockData3})} \\
\cmidrule(r){1-6}     
    $A_1$ & 5.0 & [0, 20] & $4.35^{+0.19}_{-0.19}$ & - & -\\
    $A_2$ & 12.0 &[0, 20] & $11.82^{+0.59}_{-0.55}$ & -& -\\
    $A_3$ & 2.0 & [0, 20]& $2.13^{+0.16}_{-0.16}$ & -& -\\
    $B_1$ & 100.0 &[0, 1000] & $99.93^{+0.18}_{-0.19}$ & -& -\\
    $B_2$ & 85.0 & [0, 1000]& $85.43^{+0.57}_{-0.56}$ & -& -\\
    $B_3$ & 90.0 &[0, 1000] & $89.83^{+0.16}_{-0.16}$ & -& -\\
    $s_1$ & 0.0 & [0, 10] & $0.44^{+0.34}_{-0.30}$ & -& -\\
    $s_2$ & 0.0 & [0, 10]& $3.09^{+0.66}_{-0.71}$ & -& -\\
    $s_3$ & 0.0 & [0, 10]& $0.77^{+0.30}_{-0.39}$ & -& -\\
    $\tau_2$ & 5.0 & [-50, 50]& $5.31^{+0.14}_{-0.15}$ & $5.30^{+0.27}_{-0.32}$ & $5.30^{+0.17}_{-0.15}$\\
    $\tau_3$ & 10.0 &[-50, 50] & $10.16^{+0.23}_{-0.22}$ & $10.30^{+0.45}_{-0.47}$& $10.22^{+0.24}_{-0.23}$\\
    $\Delta$ & - & [0.1, 10] &  $1.45^{+0.11}_{-0.10}$ & -& -\\      
    \hline
  \multicolumn{6}{c}{Case D: High SN with Underestimated Errors (Fig. \ref{fig:Compare})} \\
\cmidrule(r){1-6}     
    $A_1$ & 5.0 & [0, 20] & $4.965^{+0.034}_{-0.034}$ & - & -\\
    $A_2$ & 12.0 &[0, 20] & $11.787^{+0.089}_{-0.093}$ & -& -\\
    $A_3$ & 2.0 & [0, 20]& $2.003^{+0.013}_{-0.013}$ & -& -\\
    $B_1$ & 100.0 &[0, 1000] & $99.962^{+0.032}_{-0.031}$ & -& -\\
    $B_2$ & 85.0 & [0, 1000]& $85.072^{+0.085}_{-0.089}$ & -& -\\
    $B_3$ & 90.0 &[0, 1000] & $90.001^{+0.013}_{-0.013}$ & -& -\\
    $s_1$ & 0.16 & [0, 10] & $0.525^{+0.024}_{-0.023}$ & -& -\\
    $s_2$ & 0.42 & [0, 10]& $1.212^{+0.066}_{-0.063}$ & -& -\\
    $s_3$ & 0.15 & [0, 10]& $0.2478^{+0.0098}_{-0.0091}$ & -& -\\
    $\tau_2$ & 5.0 & [-50, 50]& $4.992^{+0.011}_{-0.013}$ & $4.951^{+0.056}_{-0.059}$ & $5.006^{+0.036}_{-0.012}$\\
    $\tau_3$ & 10.0 &[-50, 50] & $9.974^{+0.013}_{-0.013}$ & $9.993^{+0.041}_{-0.038}$& $9.975^{+0.042}_{-0.238}$\\
    $\Delta$ & - & [0.1, 10] &  $0.22^{+0.10}_{-0.10}$ & -& -\\    
        \hline

  \end{tabular}
\begin{tablenotes}
    \item[] Priors are uniform between the two limits given in the table.

\end{tablenotes}
  \end{threeparttable}
 \end{table}

\begin{table}
\centering
  \caption{Mock Data Results cont.}
  \label{tab:MockBlurred}
    \def\arraystretch{1.15}
    \setlength{\tabcolsep}{3pt}
    \begin{threeparttable}
    
\begin{tabular}{cccccc}

    \hline

    Parameter & Truth  & Prior  & Best Fit & ICCF & JAVELIN\\
    \hline
  \multicolumn{6}{c}{Case E: Blurred lightcurves} \\
\cmidrule(r){1-6}

    $A_1$ & 5.0 & [0, 20] & $5.170^{+0.071}_{-0.072}$ & - & -\\
    $A_2$ & 12.0 &[0, 20] & $12.114^{+0.069}_{-0.070}$ & -& -\\
    $A_3$ & 2.0 & [0, 20]& $1.574^{+0.014}_{-0.014}$ & -& -\\
    $B_1$ & 100.0 &[0, 1000] & $100.240^{+0.062}_{-0.065}$ & -& -\\
    $B_2$ & 85.0 & [0, 1000]& $85.081^{+0.063}_{-0.061}$ & -& -\\
    $B_3$ & 90.0 &[0, 1000] & $89.913^{+0.014}_{-0.014}$ & -& -\\
    $s_1$ & 0.0& [0, 10] & $1.143^{+0.046}_{-0.042}$ & -& -\\
    $s_2$ & 0.0 & [0, 10]& $0.442^{+0.057}_{-0.058}$ & -& -\\
    $s_3$ & 0.0 & [0, 10]& $0.226^{+0.013}_{-0.014}$ & -& -\\
    $\tau_2$ & 5.0 & [-50, 50]& $4.988^{+0.035}_{-0.035}$ & $4.935^{+0.052}_{-0.059}$ & $5.030^{+0.021}_{-0.022}$\\
    $\tau_3$ & 10.0 &[-50, 50] & $10.034^{+0.047}_{-0.047}$ & $9.981^{+0.093}_{-0.074}$& $9.991^{+0.049}_{-0.047}$\\
    $\Delta$ & - & [0.1, 10] &  $0.649^{+0.033}_{-0.031}$ & -& -\\
    \hline

  \end{tabular}
\begin{tablenotes}
    \item[] Priors are uniform between the two limits given in the table.

\end{tablenotes}
  \end{threeparttable}
 \end{table}

Fig.~\ref{fig:MockData} shows the model lightcurves fitted to the mock data, where the fit produced normalised residuals that look Gaussian. The top panel shows the driving lightcurve found, $X(t)$, from the running optimal average, which successfully picks up the variations in the true driving lightcurve generated from the random walk. The error envelope is small between times of 20 and 70, where there were data to calculate the shape accurately. Outwith these times the error envelope increases rapidly where there were no data to calculate $X(t)$. 

Table~\ref{tab:Testing} shows the resulting best fit parameters for this mock data set in comparison to the true values. We found that the true values for $A_2$, $A_3$, $B_1$, $B_3$ were within $1\sigma$ of the best fit parameters with $A_1$ and $B_2$ close to but outwith the error range. The time delays were recovered successfully, with $\tau_2$ within $1\sigma$ and $\tau_3$ very slightly outwith the error range. As the time delays are the parameters of interest, the accuracy of their error bars are investigated further in Sec. \ref{sec:ErrorBars}.

\begin{figure*}
\hspace*{-1.0cm}                                                           
    \centering
	\includegraphics[width=17cm]{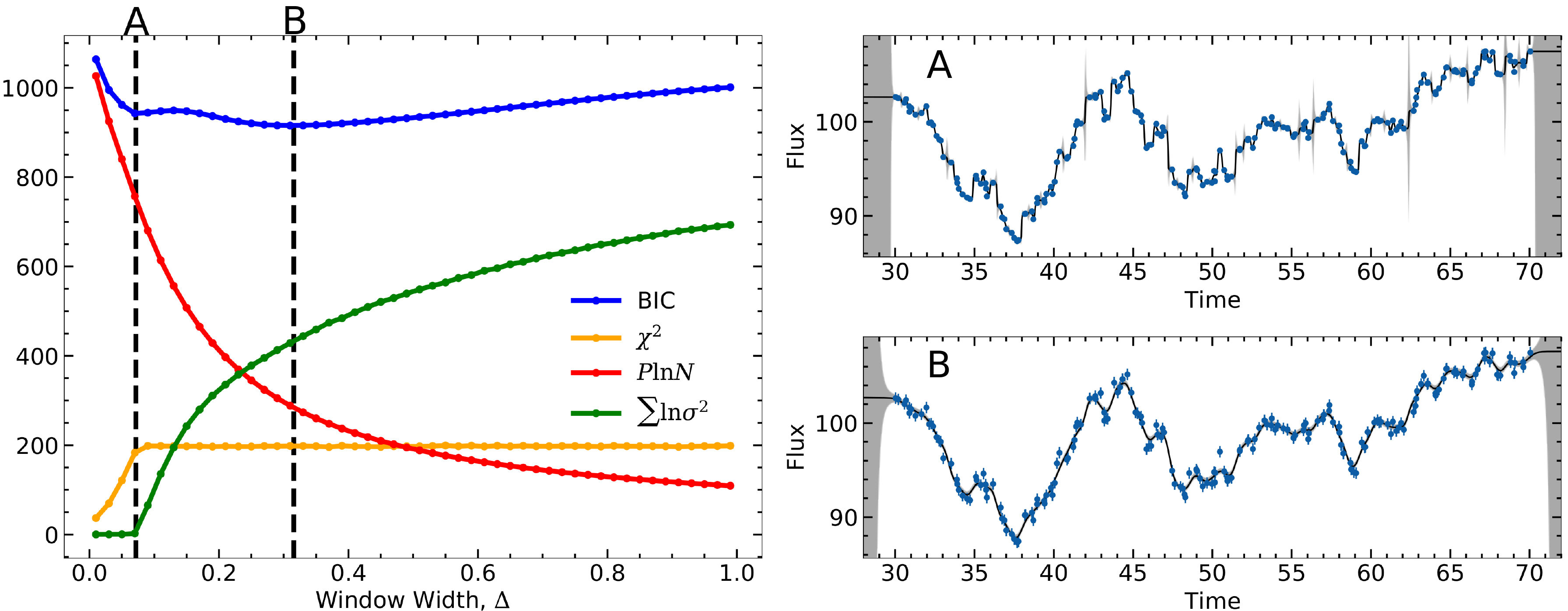}
    \caption{{\it Left:} Plot of the BIC and its constituent components as a function of window width, $\Delta$ for a Gaussian window function. The vertical dashed lines show the best fit value of $\Delta$ where the extra variance parameter is not included (A) and is included (B). {\it Right:} The fit of the model to the single lightcurve where the extra variance parameter is not included (A) and is included (B), corresponding to the dashed lines in the left panel.}
    \label{fig:BICPlot}
\end{figure*}

Interestingly, extra errors were added by the model, even though they were not deliberately underestimated when generating the mock data set. To understand this, we explored how the BIC and it components vary as a function of the window width, $\Delta$. To do this, the model was fitted to a single lightcurve numerous times, each at a different fixed value of $\Delta$. This is shown in Fig.~\ref{fig:BICPlot}. 

Initially when $\Delta$ is small, the ROA is very flexible and thus the penalty that scales with the number of parameters, $P\ln N$, is very large (red). This balances with the very low $\chi^2$ to create a minimum in the BIC, labelled A on the figure. At this point no extra errors are added, as the model is flexible enough to pick up all the variations, therefore the penalty for adding this additional error is zero (green). If the model is fitted without including a parameter that adds extra errors, the best fit solution is A, which corresponds to this minimum in the BIC. 

As $\Delta$ increases beyond this point, the BIC begins to rise again as extra errors are now being added by the model to accommodate the smoother ROA, increasing the penalty for adding additional errors (green). The smoother ROA has less effective number of parameters, so this penalty (red) is no longer dominant, and so there is another minima in the BIC, labelled B. This is a lower minimum and so it corresponds to the best fit model where extra errors are added as a free parameter. These two solutions can be seen in the right panels of Fig.~\ref{fig:BICPlot}.

This effect means that out algorithm is over cautious when including the noise model, by increasing the flux errors, increasing the uncertainty in the time delay. This effect is safe as it does not cause any bias in the results and just makes the results less certain by sacrificing the fastest variations for noise. As explained previously the higher the extra variance parameter, the wider the width of the window function, resulting in a smoother model. This smoother model therefore measures time delays using less information resulting in larger uncertainty in the time delay.

\begin{figure*}
\hspace*{-0.2cm}                                                           
    \centering
	\includegraphics[width=18cm]{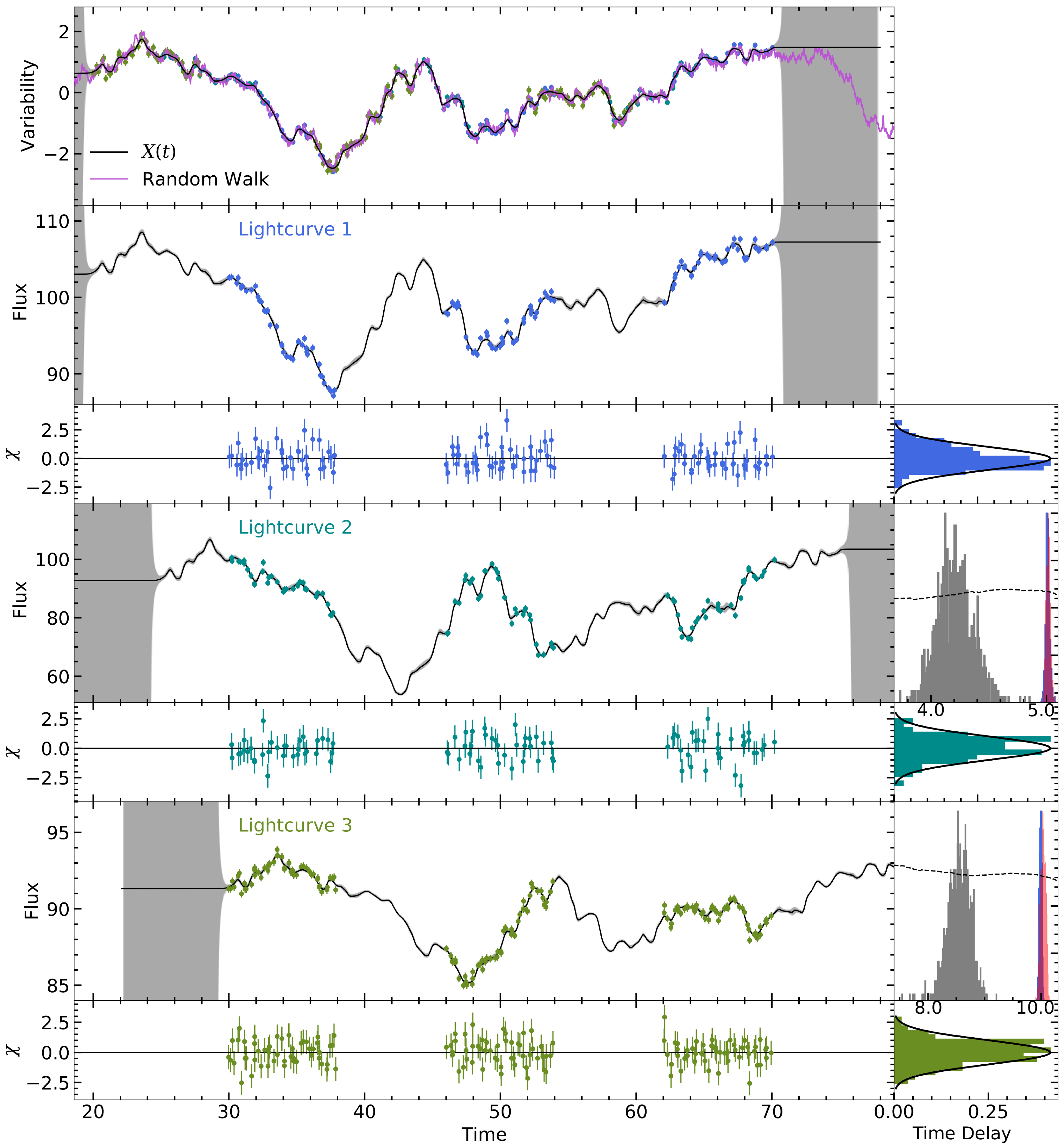}
    \caption{The same as Fig.~\ref{fig:MockData} but for Case B where large gaps are inserted into the lightcurves. Additionally the top panel shows the data after the shifting and stacking stage of the fitting procedure (Section~\ref{sec:Fitting}), from which $X(t)$ is calculated. The signal to noise ratio (variability to noise) is $\sim 20$}
    \label{fig:MockData2}
\end{figure*}

The benchmark data are well sampled with a high signal-to-noise. We therefore present three further mock data sets, one with large gaps inserted (Case B),  one with a signal to noise lower by a factor of 10, of $\sim$ 2  (Case C), and finally one with lightcurves with different amounts of blurring  (Case E). For this we used the same random walk lightcurve as the previous mock data set, with the same true values of the parameters. The first of these (Case B) can be seen in Fig.\ref{fig:MockData2}, where two large gaps were inserted into the lightcurves between times of 38 to 46 and 54 to 62. This is a common scenario for ground based observing campaigns where there are yearly gaps due to the object being too close to the sun on the sky. We find that the delays are successfully recovered with $\tau_2 = 5.010^{+0.018}_{-0.013}$ and $\tau_3 = 9.993^{+0.021}_{-0.024}$, which contain the true values of $5$ and $10$. This case demonstrates one of the major strengths of our method as using all of the available data to calculate $X(t)$, provides information within the gaps of the individual lightcurves. This is demonstrated in the top panel of Fig.\ref{fig:MockData2}, where the data used to calculate the ROA is shown, after shifting and stacking as described in Section \ref{sec:Fitting}. Even if no data are available, the error envelope of the ROA will expand accordingly as it interpolates across a gap.

The next simulation (Case C) is where the lightcurves are sampled with a lower cadence and a lower signal to noise of $\sim$ 2 . This provides a more typical case for larger surveys such as the The Sloan Digital Sky Survey Reverberation Mapping (SDSS-RM) project \citep[][]{Shen2015} and upcoming surveys such as the Legacy Survey of Space and Time (LSST) at the Vera C. Rubin Observatory.

\begin{figure*}
\hspace*{-0.2cm}                                                           
    \centering
	\includegraphics[width=18cm]{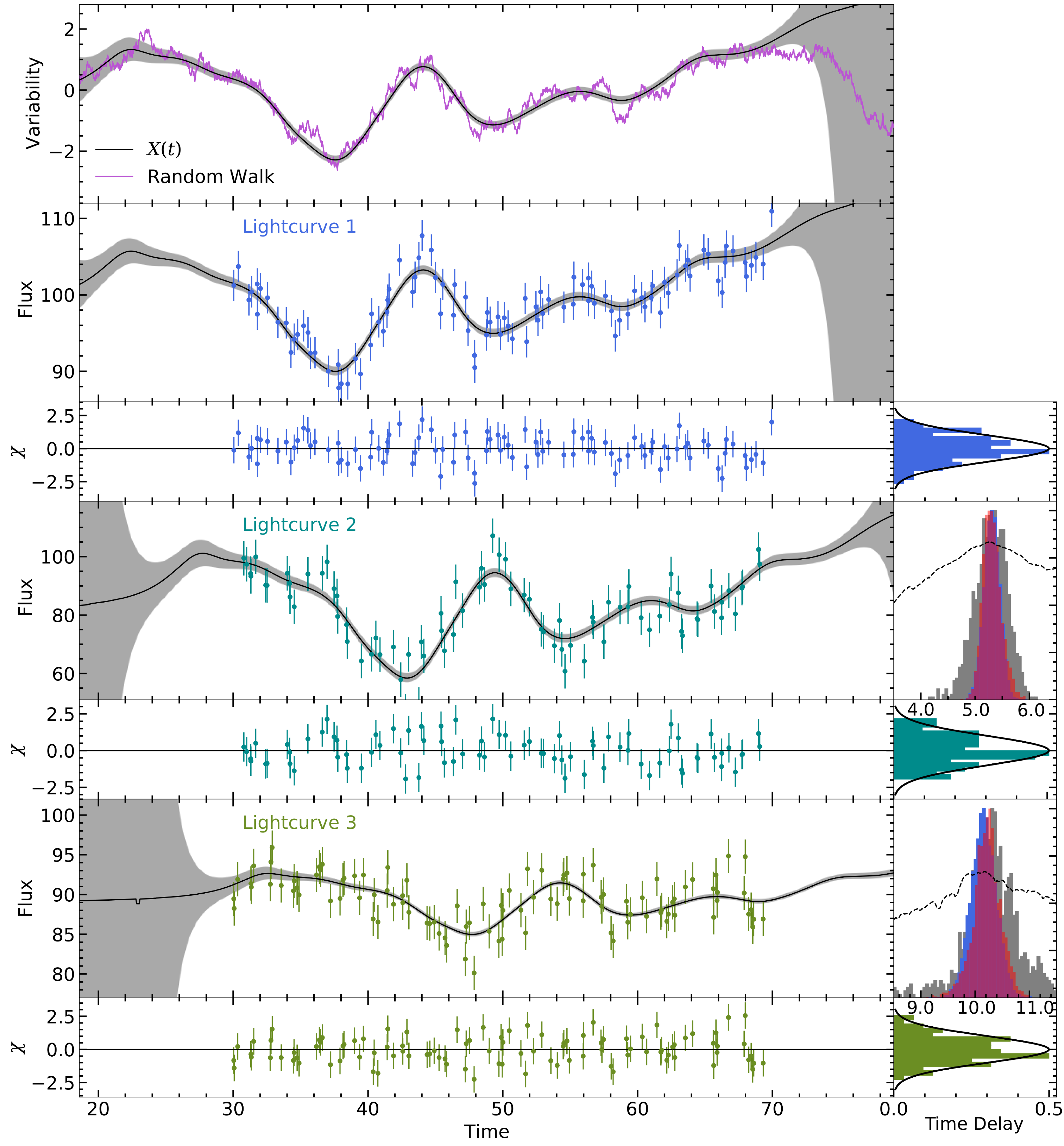}
    \caption{ The same as Fig.~\ref{fig:MockData} but for Case C, which contains a lower cadence and a lower signal to noise ratio of $\sim$ 2.}
    \label{fig:MockData3}
\end{figure*}

The number of epochs in the first lightcurve is lowered to 100, the second to 80 and the third to 100. This is shown in Fig.~\ref{fig:MockData3}, where the noisier data results in a smoother model with a lower window width, $\Delta$, and a wider error envelope due to the poorer data.  We find time delays of $\tau_2 = 5.31^{+0.14}_{-0.15}$ and $\tau_3 = 10.16^{+0.23}_{-0.22}$, where the true value for $\tau_3$ is within $1 \sigma$ whereas the true value for $\tau_2$ is $\sim 2 \sigma$ from the measured value.

The final simulation (Case E) is where lightcurves 2 and 3 are blurred by different amounts, in addition to being shifted in time. We do this by convolving the random walk with a Gaussian with widths of 1 and 2 for lightcurves 2 and 3 respectively. This simulates the convolution of the driving lightcurve with a response function, which is a strong effect for BLR RM, where emission line lightcurves are smoothed relative to the continuum. PyROA assumes the same shape for each lightcurve, with a single level of smoothing given by the window width $\Delta$. The results for this test case are shown in table \ref{tab:MockBlurred}, where the true time delays are recovered accurately. By assuming a single level of smoothing, the resulting value of $\Delta$ is somewhere between the most smoothed lightcurve (3) and the most flexible lightcurve (1), where the error bars of these two lightcurves are expanded to be consistent with the single ROA. Despite this PyROA is still able to recover the mean delay although the expanded error bars increase the uncertainty. Allowing a different $\Delta$ for each model lightcurve would account for a Gaussian transfer function, however this is a symmetric transfer function. Theoretical transfer functions for accretion disk reverberation \citep[e.g.][]{Starkey2016} are asymmetric which if not accounted for, can cause a bias towards a small mean delay \citep[][]{Chan2020}. This is a problem present in JAVELIN, which uses a uniform transfer function and also a problem for ICCF, which treats the time delays symmetrically. In a future paper we extend PyROA to use an asymmetric transfer function with respect to the mean delay. A plot of the PyROA fit for Case E can be found in the supplementary material in Fig. 34.  

To place these results in context, we compared them to two popular methods for measuring time delays, ICCF \citep{ICCF} and JAVELIN \citep{JAVELIN}.
\subsection{Comparison to ICCF and JAVELIN}
\label{sec:Compare}

To compare to the interpolation cross-correlation method (ICCF) \citep{ICCF}, we used the code PyCCF \footnote{\url{https://bitbucket.org/cgrier/python_ccf_code/src/master/}} \citep{PyCCF}. We used an interpolation grid between 0 and 15 with a spacing of 0.01. To sample errors for the time delays, this code uses flux randomisation/random subset selection (FR/RSS) method \citep{Peterson1998}, which measures the lags from many realisations of the CCF. The measured delays are based on the centroid of the CCF using values of $r > 0.8\,r_{\textrm{max}}$, where $r_{\textrm{max}}$ is the maximum value of the CCF. 

We also compared our ROA algorithm to another popular method for measuring time delays, JAVELIN \footnote{\url{https://github.com/legolason/javelin-1}} \citep{JAVELIN}.  This method uses a damped random walk to model the variability, which is first determined from a reference lightcurve and then subsequently shifted and blurred to fit the other lightcurves. To fit to our mock data, we used the first lightcurve as a reference.

\begin{figure}
\hspace*{-0.2cm}                                                           
    \centering
	\includegraphics[width=\columnwidth]{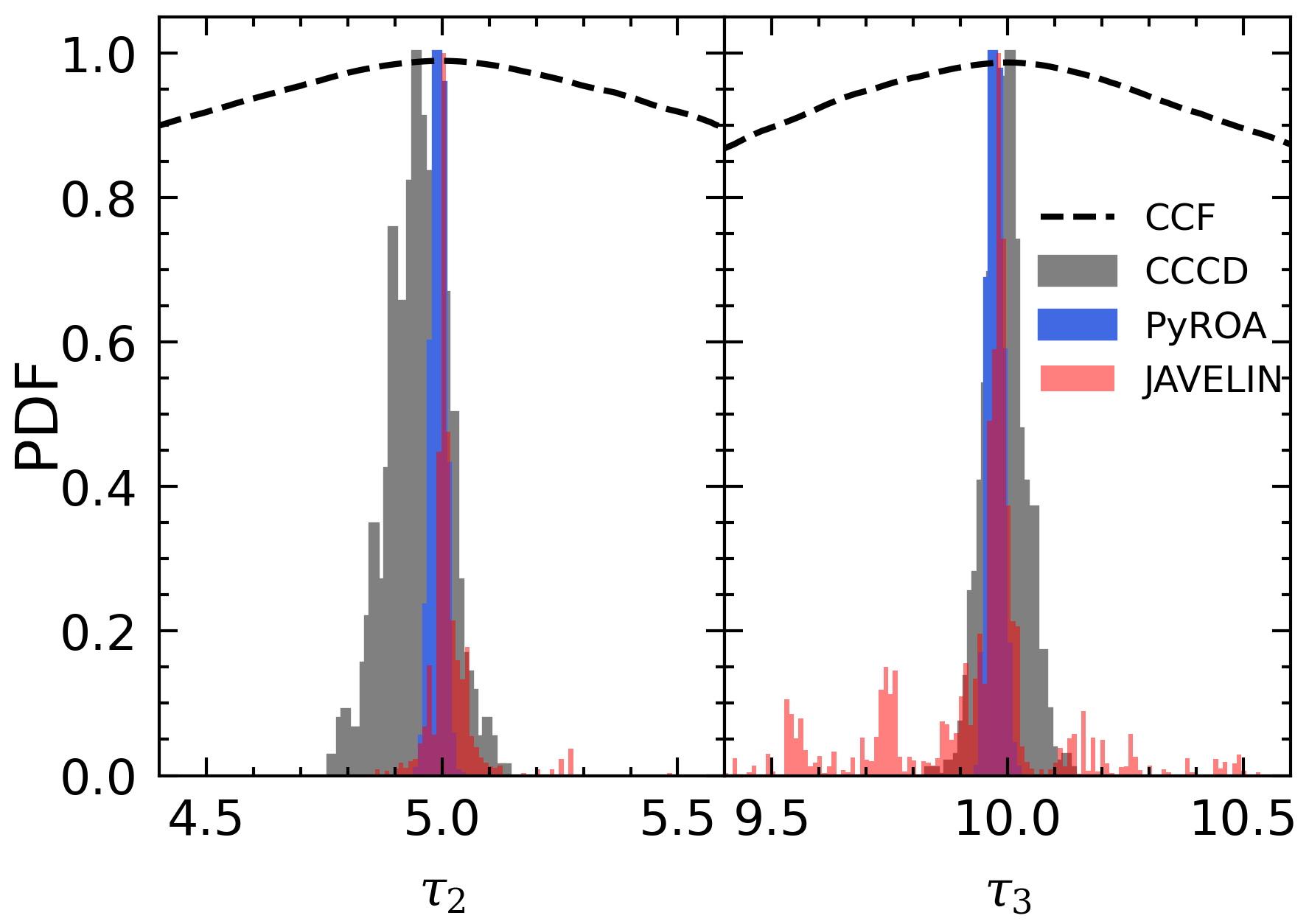}
    \caption{Posterior probability distributions of the time delays between the high S/N mock lightcurves but where the flux errors are deliberately underestimated by a factor of 5. The cross-correlation centroid distribution is shown in gray, our method is shown in blue and JAVELIN is shown in red. The dashed line shows the cross-correlation function.}
    \label{fig:Compare}
\end{figure}

This method consistently finds smaller uncertainties than the ICCF \citep{Yu2020}, however it does not account for poorly estimated errors on the flux measurements of AGN lightcurves, making it sensitive to this effect. We applied both of these methods to the three mock data sets discussed previously. For the first data set, the ICCF finds significantly larger uncertainties than our method whereas JAVELIN on average finds slightly smaller uncertainties. These are shown in Table~\ref{tab:Testing} and the posterior probability distributions are compared to our method in the right panels of Fig.~\ref{fig:MockData}. One reason for JAVELIN finding smaller uncertainties may be that it assumes the error bars on the flux data are accurate whereas our method expands these to account for poorly estimated errors. To investigate this further we fit the model again but with error bars on the flux measurements that were deliberately underestimated by a factor of five (Case D). The resulting probability distributions for the time delays are shown in Fig.~\ref{fig:Compare}, comparing our method to the cross-correlation and JAVELIN. The best fit parameters are shown in table \ref{tab:Testing}. Our method measured consistent delays as previously found whereas JAVELIN had some difficulties due to over-fitting the noise. While the first delay was measured successfully by JAVELIN, finding $\tau_2 = 5.006^{+0.036}_{-0.012}$, its errors were larger and asymmetrical. The second delay had similar problems, with a measurement of $\tau_3 = 9.975^{+0.012}_{-0.24}$, where several smaller peaks in the probability distribution skewed the lower error estimated, resulting in very uneven error bars. The cross-correlation is similar here to our method, successfully recovering the delays similar to before with similarly large uncertainties. This shows the importance of including a noise model when there is no prior knowledge that the flux measurement errors are accurately known.

We find that the extra error parameters increase by larger than the amount it was deliberately underestimated. This is similar to what we found when the uncertainties were not underestimated, that the algorithm is over cautious, sacrificing the very fastest variations as noise. This makes it robust when dealing with data such as this test case, where the flux errors are underestimated, and can still recover an accurate time delay. This is tested more thoroughly in section \ref{sec:ErrorBars}. We also note that the error bars were not expanded as much for lightcurve 2 than the case where the uncertainties were not underestimated. We suspect this is due to the scatter in the size of the flux error bars across lightcurve 2 due to the uniform random number we added when generating the data, as described in section \ref{sec:Mock}. This scatter was larger for lightcurve 2 and therefore the extra variance added to the whole lightcurve by the algorithm is higher to make the data consistent with the other lightcurves when calculating the ROA. In the underestimated case, the division by a factor of 5 reduces the strength of this scatter and therefore the flux errors are not expanded as much to be consistent. 

For the second data set, shown in Fig.~\ref{fig:MockData2}, the large gaps proved difficult for the ICCF, causing both delays to be massively underestimated. This is likely due to the linear interpolation across the gaps which are being treated as a feature in the lightcurve when measuring the cross-correlation function. This is a known problem with the ICCF. In comparison, our method was successful in recovering the true delays as was JAVELIN. In this case, our method produced uncertainties comparable to JAVELIN.

The  third data set contained few points, with a lower signal to noise ratio of $\sim$ 2. Our method is consistent with ICCF and JAVELIN and finds uncertainties comparable to JAVELIN. Interestingly, the uncertainty in the delays from the ICCF are closer to our method and JAVELN than in the high signal to noise case. 

The final data set was where the 2\textsuperscript{nd} and 3\textsuperscript{rd} lightcurves were blurred by a different amount. Comparing to ICCF and JAVELIN we see a similar result to the previous case, where the ICCF errors are the largest and PyROA is reasonably similar to JAVELIN in its error estimates.

\subsection{Verifying the Accuracy of the Error Bars}
\label{sec:ErrorBars}
\begin{figure*}
    \centering
	\includegraphics[width=14cm]{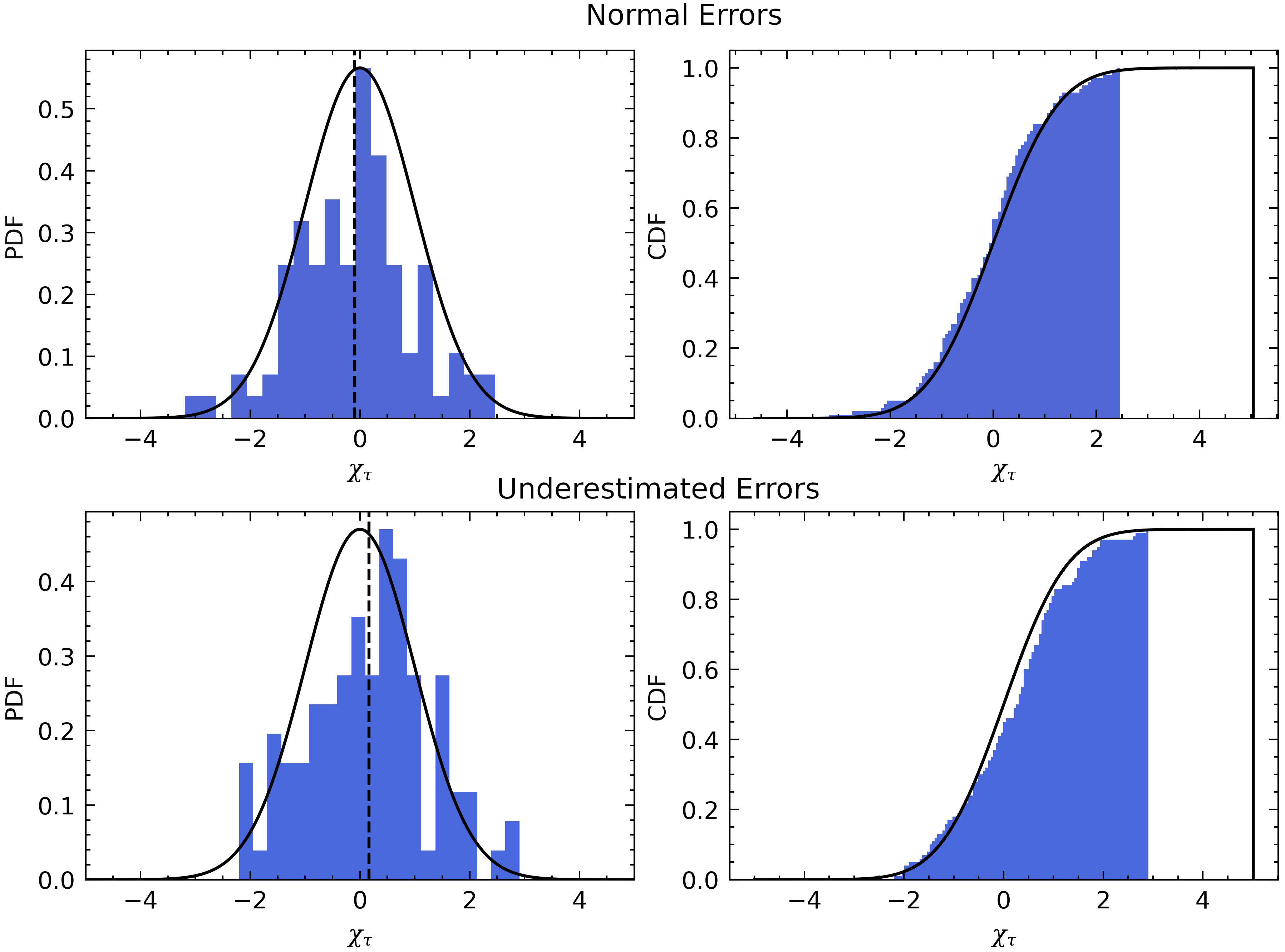}
    \caption{Results of the error bar testing for the time delay parameters. The top panel shows the results for normal flux errors whereas the bottom panel shown the testing where the flux errors were deliberately underestimated by a factor of 5. {\it Left:} Probability distribution of the normalised residuals of the pairs of time delay parameters for 50 random walk lightcurves, relative to the true value (blue), with a Gaussian distribution with a mean of zero and standard deviation of 1 in black. The black dashed line shows a sample mean of -0.09 in the top panel and a sample mean of 0.16 in the bottom panel. {\it Right:} Cumulative distribution of the normalised residuals (blue) compared to the Gaussian (black).}
    \label{fig:Resid}
\end{figure*}

To ensure that the errors in the time delays are being predicted accurately, 50 additional mock data sets were fitted where each of these data sets were based on a different random walk. We tested the robustness of the errors by calculating the normalised residuals of the time delay parameters with respect to the true value. This was calculated by subtracting the true value from the measured value and dividing by the average of the errors. This creates a distribution which should have a mean of zero and a standard deviation of one for properly estimated error bars. As there are 2 time delay parameters per data set, there is a total of 100 samples. The resulting probability distributions and cumulative distributions for all 100 samples are shown in the top panel of Fig.\ref{fig:Resid}. 

This distribution has a sample mean of $-0.09 \pm 0.11$ and a sample standard deviation of $1.06 \pm 0.16$ which is consistent with the expected result for well defined error bars. A mean close to zero suggests no systematic bias in the method while a variance of one suggests the error bars are accurately estimated. Furthermore we performed a Kolmogorov-Smirnov (K-S) test \citep{KSTest}, which tests whether the distribution of our samples is drawn from an underlying Gaussian distribution that has a mean of zero and variance of one. This is the null-hypothesis - that both distributions are the same - which typically requires a p-value $< 0.05$ to reject.We use the one-sample KS test from the {\sc scipy.stats} package. This calculates a p-value based on the maximum distance between the measured cumulative distribution function (CDF) and the expected normal CDF as well as the sample size.  We find a p-value of $0.67$, which suggests that our samples are consistent with the expected normal distribution. These tests confirm that our method is producing accurate error bars.

As discussed in section \ref{sec:Compare}, a major advantage of PyROA is the inclusion of a noise model that can account for underestimated flux errors. To verify that the true time delays can be recovered consistently by PyROA in this case, we repeated the previous test, fitting to 50 mock data sets but this time where the flux errors are underestimated by a factor of 5. We again calculated the normalised residuals of the time delay parameters which are plotted in the bottom panel of Fig.\ref{fig:Resid}. We find a sample mean of $0.16 \pm 0.11$ and a sample standard deviation of $1.12 \pm 0.18$. The standard deviation is consistent with 1 suggesting that the size of the error bars are accurate however the mean is $>1\sigma$ from zero, although it is close at $1.45\sigma$. A KS test returns a p-value of 0.06 which is greater than the rejection criterion of $>0.05$, and therefore provides weak evidence that our samples are drawn from a normal distribution. The low p-value is driven largely by the mean and so shifting the distribution by the mean finds a p-value of 0.7 which is strongly consistent with a normal distribution. We suspect that the mean of $0.16 \pm 0.11$ is due to a lack of samples and not a systematic bias as it is only $1.45\sigma$ from the true value. This would suggest that PyROA is able to obtain accurate time delays where the flux errors are underestimated by a factor of 5, albeit with weaker evidence than the normal flux error case.

\subsection{Choice of Window Function}
\begin{figure*}
    \centering
	\includegraphics[width=17cm]{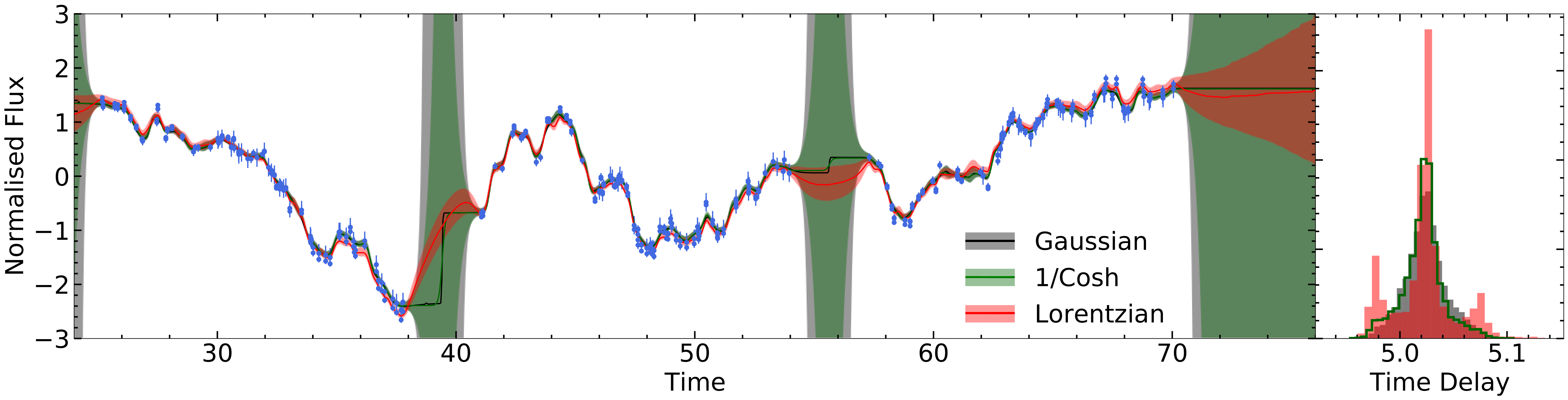} 
    \caption{Calculation of the running optimal average at the stacking stage of the fitting procedure (Section~\ref{sec:Fitting}), when fitting two lightcurves that contain large gaps. Three different window functions were used: a Gaussian shown in black, inverse-cosh shown in green and a Lorentzian in red. The error envelopes are shown as the shaded colour. The right panel shows the probability distribution for the time delay between the lightcurves for each of the window functions.}
    \label{fig:WindowTest}
\end{figure*}

For all of the testing so far, a Gaussian window function has been used. As many choices are possible, we tested the effect this has on the results of fitting the model to mock data. We used two lightcurves that included large gaps as the window function has a large effect on the error envelope calculated from equation (\ref{eqn:ROAerrs}). Fig.~\ref{fig:WindowTest} shows the calculation of the driving lightcurve using three different window functions, Gaussian, inverse-cosh, and Lorentzian, given by equations (\ref{eqn:Gauss}, \ref{eqn:1/cosh}, \ref{eqn:Lorz}) respectively. The right panel of this figure shows the resulting posterior distributions for the time delay between the two lightcurves. 

The main difference between them is how the error envelope behaves when there is a lack of data. The error envelope of the Gaussian window function rapidly increases in the gap whereas the error of the inverse-cosh increases slower but still becomes very large in the gaps. This is largely due to the wider wings of the inverse-cosh function compared to the Gaussian, which will allow data points far from centre of the window to still slightly contribute to the running optimal average. This effect is even more pronounced when using a Lorentzian window function, which has even wider wings. Here the error envelope only increases slightly within the gaps and increasing slower of the ends of the data. At the other extreme, a boxcar function drops to zero outwith the window, which would result in an error envelope becoming infinity within these gaps.

The probability distributions of the time delays are of similar width and peak for each window function however the Lorentzian shows an unusual feature, where there are two smaller peaks to each side of the main peak. The physical cause of this is unknown and as there is no real difference in the probability distributions for the Gaussian and inverse-cosh functions, we continue using a Gaussian window function for the remainder of this paper.

\section{Gravitational Lensing Time Delays}
\label{sec:GravLensing}
In order to test this method with real data, we applied it to quasars that are gravitationally lensed by an intermediate object such as a galaxy or galaxy cluster. The lens causes the light from the quasar to form multiple images of itself on the sky. As the light from the quasar travels along vastly different paths to form each image, a time delay is induced, due to the geometric difference in the path length that the light travels along \citep{Cooke&Kantowski1975}, and the different gravitational potentials that the photons experience \citep{Shapiro1964}. As the lensed quasar is variable, obtaining lightcurves for each of these images allows the time delay between images to be measured, using the method outlined in this paper. There are also microlensing effects taking place, which causes the brightness of the image to vary slowly with time due to objects in the lens moving relative to the source and observer \citep{Refsdal1964}. These effects can also be modelled by including a slow varying component to the model, that modulates the brightness of each image relative to some reference image. Accounting for these effects are important for obtaining accurate delay measurements as they can create extraneous features in the lightcurves, resulting in poorly estimated delays.

We used public data from the COSmological MOnitoring of GRAvItational Lenses (COSMOGRAIL) project \citep[][and references therein]{COSMOGRAIL}, to measure the time delays between gravitationally lensed images of 34 different quasars and model microlensing effects. This requires a slightly different model than previously used, specifically to model the microlensing effects. We model the flux of each image, indexed $i$, where $i=1$ is the image which the magnifications are measured relative to. Therefore the model flux is given by

\begin{equation}
\label{eqn:lensModel}    
    f_{i}(t)= \left.
    \begin{cases} 
      A_{1}X\left(t \right) + B_{1} & i=1 \\
      \left[ A_{1}X\left(t -\tau_i \right) + B_{1} \right] 10^{-0.4 P_i(t)} & i>1 \\
   \end{cases} 
   \right\} \ ,
\end{equation}
where $A_{1}$ represents the rms flux of the reference image, $B_{1}$ represents the mean flux of the reference image, $\tau_{i}$ represents the time delay of image $i$ relative to the first image and $X(t)$ is the driving lightcurve, normalised such that $\left< X\right>_t=0, \quad \left< X^2 \right>_t=1$. The magnitude different due to microlensing is given by a 4\textsuperscript{th} order polynomial, $P_i(t)$. This is given by 

\begin{equation}
    P_i(t) = \sum_{j=0}^{4} P_{i, j}\, \eta^j(t),
\end{equation}
where the argument, $\eta(t)$, is time normalised such that it runs between -1 and 1 over the range of the data. This is calculated by 
\begin{equation}
\label{eqn:eta}
    \eta(t) = \frac{2\left(t - t_0\right)}{\Delta t},
\end{equation}
where $t_0$ is the midpoint time of the data and $\Delta t$ is the length of time between the start of observing and the end. Normalising the time in this manner ensures that the ``pivot point'' of the polynomial is at the centre of the data.

As we model the flux of the lightcurves, the data were converted from magnitudes into arbitrary flux units before modelling. This was done using 
\begin{equation}
    f(t) = 3.0128 \times 10^{-5} 10^{-0.4 m(t)},
\end{equation}
where $m(t)$ is the magnitude of the image as a function of time, $t$. This conversion gives the flux in arbitrary units with magnitudes on the order of unity for the majority of the COSMOGRAIL data. For DES 2325-5229, HE 0047-1756, PDJ 1606-233, SDSS J1515+1511 and WG 0214-2105, this factor was $3.0128 \times 10^{7}$, as the data here measured magnitude relative to some other arbitrary value. As the parameters $A_1$ and $B_1$ have flux units, ensuring that they are not a drastically different order of magnitude (e.g. 14 orders different) to the other parameters helps when initialising the walkers to keep them linearly independent. 
 
The method used by COSMOGRAIL measures delays between every image rather than to a single reference image \citep{Tewes2013}. Therefore, to be directly comparable, we fit our model numerous times where the time delays are measured relative to a different image each time e.g. if there are three images (A, B, C), the model is fitted firstly with image A as the reference and then with image B as the reference. This allows all the inter-image delays to be obtained as well as the relative microlensing between all the images.

We used uniform priors between sensible limits for all the parameters and used 15~000 samples discarding the first 10~000 as burn-in for the two lightcurve data. For objects with three/four lightcurves we used 20~000 samples with 15~000 as burn-in. For the HE~0435-1223 we used 35~000 samples and a burn-in of 30~000 as the extra parameters required a long burn-in in this case. The number of walkers for each is $\gtrapprox$~twice the number of sampled parameters. Specifically this is $7N_l + 3$, where there are $N_l$ lightcurves, although this number was chosen fairly arbitrary meeting the only requirement of being $>$ twice the number of sampled parameters. Here we also include the noise model where there is extra variance added to the flux errors of each lightcurve as described previously.

\subsection{Results} 

We show our results for the publicly available COSMOGRAIL lightcurves in Table~\ref{tab:cosmograil}. As our method measures the time delay in the opposite direction, we take the negative of the measured time delay posterior distributions in order to compare directly with COSMOGRAIL i.e $\tau_{AB} = - \tau_2$ where $i=1$ and $i=2$ represent images A and B respectively, using equation (\ref{eqn:lensModel}). The table also shows the coefficients of the 4\textsuperscript{th} order polynomial used to model the microlensing variability. The 0\textsuperscript{th} order coefficient is of particular interest as this provides the mean difference in magnitude between the images. The higher order terms describe how the magnitude varies with time around this mean.

Comparing to previous analysis, our results find time delays with consistently smaller uncertainties than the COSMOGRAIL analysis. For most of these objects, the error regions overlap as we find delays that are consistent, however a few objects show some interesting results. Firstly, some show significantly smaller errors such as DES J0408-5354, SDSS J0832+0404 and DES 2325-5229. Secondly, we also find delays for numerous objects where previously delays were not able to be measured, e.g. SDSS J1226-0006, SDSS J1320+1644 etc. Naturally these objects have large uncertainties but are somewhat constrained. Lastly, we find some disagreement with the previous analysis for a few objects. In particular HE 2149-2745 and HS 0818+1227 show strong disagreement, however \citet{Millon2020a} notes that they are tentative in their estimate for these two objects, so some disagreement is not entirely unexpected. UM 673 showed strong disagreement with \citet{Millon2020a}, and although they are again uncertain in their measurement, with other studies finding $\tau_{AB} = -72 \pm 22$ days, \citep[][]{Oscoz2013} and $\tau_{AB} =-95^{+5}_{-16}$ days \citep[][]{Koptelova2012}, these overlap with their result while ours is $\sim 7\sigma$ further from these results. Our delay is likely inaccurate as this object shows little intrinsic variability as shown in Fig.~\ref{fig:UM673}, meaning that the microlensing variability may distort and prevent an accurate delay from being obtained. 

We included the extra error parameters on the fluxes for each lightcurve to account for underestimated errors. We find that the error bars of fluxes were expanded for these fits, for some more than others. For example in the case of HE 0435-1223 discussed later, the long observation time means that the microlensing curve is unable to pick up all the variations, resulting in the error bars being expanded more than DES J0408-5354 for example, where the microlensing variations are small. This results in the time delays being less certain, however they are all more certain than the previous COSMOGRAIL results so we present our results including the extra variance. This therefore means that more precise delays are possible with our method, if we did not use the extra variance parameters.

A full list of plots can be found online\footnote{\url{https://dx.doi.org/10.5281/zenodo.5060008}}, however we show a few cases representative of the full sample. These plots show our best fit model overlaid on the lightcurves, with the lower panel of each image showing the microlensing behaviour. This is the magnitude difference relative to image A, calculated by dividing the fluxes of the other images by the model for image A shifted to remove the time delay. Specifically this is:
\begin{equation}
\label{eqn:deltam}
    \Delta m = -2.5 \log\left[ \frac{f_i(t)}{ A_{1}X\left(t -\tau_i\right) + B_{1}} \right], 
\end{equation}
where $f_i(t)$ are the flux data for lightcurve $i$. This is the component modelled by the low order polynomial in equation (\ref{eqn:lensModel}), which is shown as a black solid line, overlaid on top of this data.

\begin{figure*}
\hspace*{-0.8cm} 
    \centering
	\includegraphics[width=17cm]{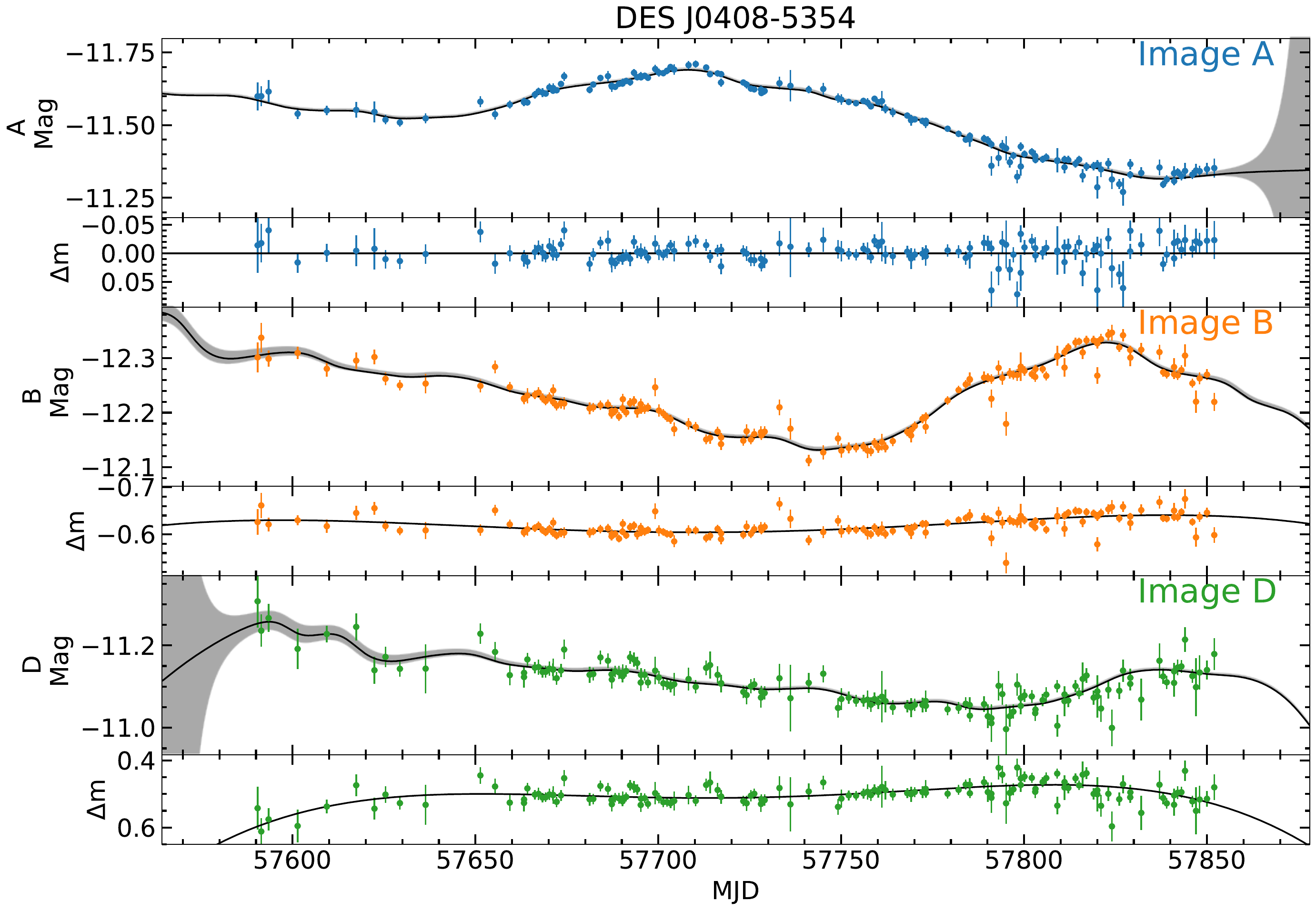}
    \caption{ Lightcurves for DES J0408-5354, overlaid with our best fit model in black, with the grey shaded region showing the error envelope in the ROA. The color indicates the image, with the lower panels of each image showing the microlensing behaviour relative to image A. The data points for this are calculated from equation (\ref{eqn:deltam}), and are represented by a 4\textsuperscript{th} order polynomial in the model, shown in black.}
    \label{fig:DES J0408}
\end{figure*}

\begin{figure*}
\hspace*{-0.8cm} 
    \centering
	\includegraphics[width=17cm]{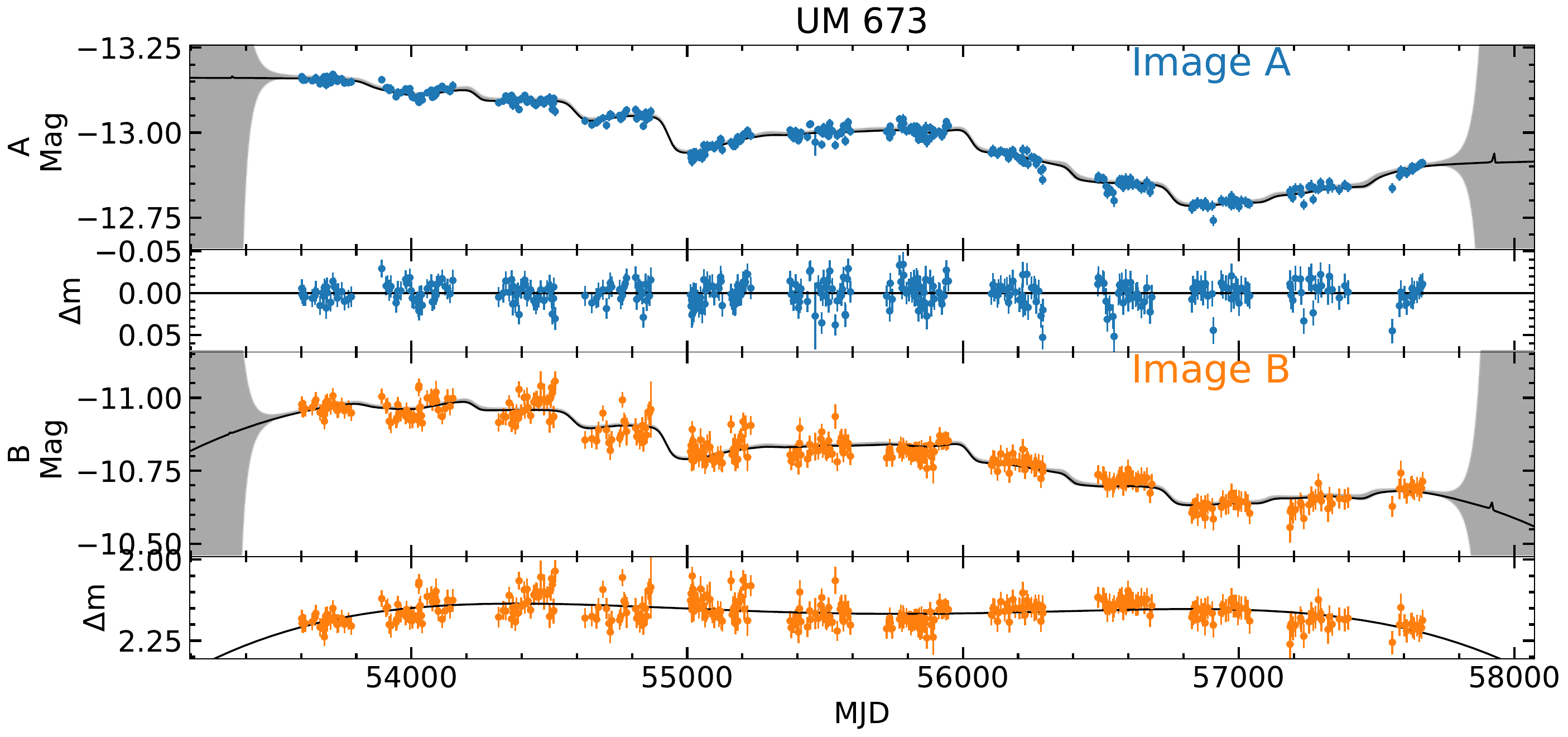}
    \caption{ Same as Fig.~\ref{fig:DES J0408} but for UM 673.}
    \label{fig:UM673}
\end{figure*}

Fig.~\ref{fig:DES J0408} shows the results for DES J0408-5354, a quadruply-imaged quasar at $z=2.375$, lensed by a galaxy at $z=0.597$ \citep{Lin2017}. Three of these images, A, B, D, were observed over 7 months with the MPIA 2.2m telescope and 1.2m Euler Swiss telescope at La Silla \citep{Courbin2018}, yielding three lightcurves. This object is particularly noteworthy as we found significantly smaller errors for the time delays in comparison with the previous analysis from \citet{Courbin2018}.

\begin{figure*}
\hspace*{-0.8cm} 
    \centering
	\includegraphics[width=17cm]{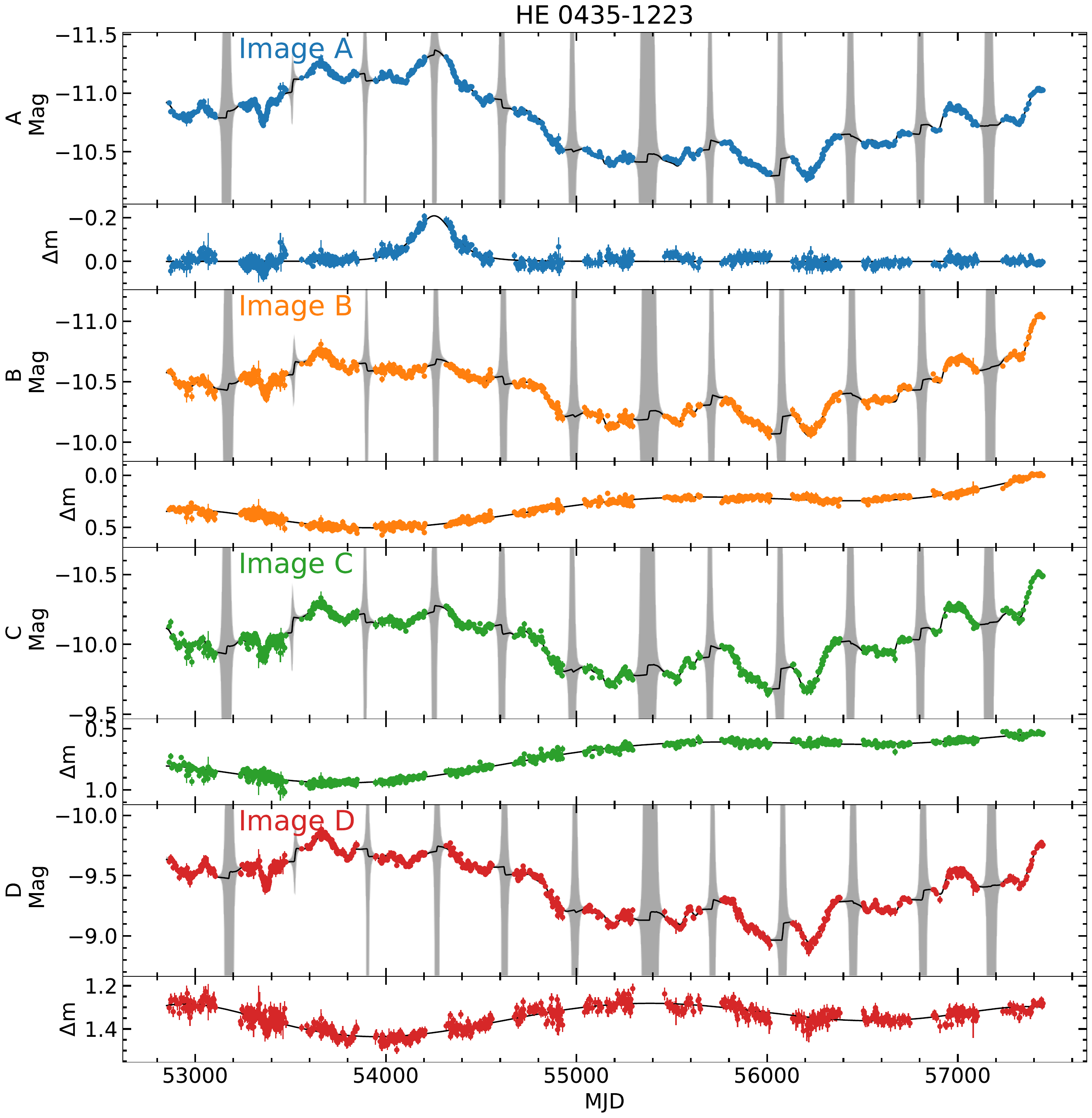}
    \caption{ Lightcurves for HE 0435-122, overlaid with our best fit model in black, with the grey shaded region showing the error envelope in the ROA. The color indicates the image, with the lower panels of each image showing the microlensing behaviour relative to image A. The data points for this are calculated from equation (\ref{eqn:deltam}), and are represented by a 6\textsuperscript{th} order polynomial in the model, shown in black. For image A, magnification for lensing due to a point mass is inserted.}
    \label{fig:HE 0435}
\end{figure*}

Another interesting object was HE 0435-1223, which is a quadruply-imaged quasar at $z_S=1.693$, with a lens at at $z_L=0.454$ \citep[][]{Wisotzki2002}. All four images were monitored over 13 years, providing 4 lightcurves \citep[][]{Millon2020a}. Fig.~\ref{fig:HE 0435} shows the results for this object. This object contained much more data than the other objects and thus it was difficulty to achieve a good fit with only a 4\textsuperscript{th} order polynomial to model the microlensing, so we used a 6\textsuperscript{th} order polynomial. We also noticed that there is a sharp increase in the brightness of image A towards its peak at a MJD~$\sim54250$, that was unable to be fitted with a simple polynomial. Therefore we inserted an extra magnification term for image A, that of a lensing due to a point mass \citep[][]{Paczynski1986}. This magnification as a function of time, $t$, is defined as
\begin{equation}
    A\left[u(t)\right] = \frac{u^2 +2}{u \sqrt{u^2 +4}}, \quad u(t)=\sqrt{\left( \frac{t - t_0}{t_E}\right)^2 + u_0^2}, 
\end{equation}
where we add three new parameters to the model: the time of maximum magnification, $t_0$, the Einstein ring radius crossing time in days, $t_E$ and the impact parameter, $u_0$, in units of the number of Einstein ring radii. We restricted the prior of $t_0$ to be uniform between 54200 to 54300 and $t_E$ uniform between 10 and 200 days, to ensure that this fitted the sharp peak in image A. This means the model becomes
\begin{equation}
\label{eqn:lensModel2}    
    f_{i}(t)= \left.
    \begin{cases} 
      \left[ A_{1}X\left(t \right) + B_{1}\right] A & i=1 \\
      \left[ A_{1}X\left(t -\tau_i \right) + B_{1} \right] 10^{-0.4 P_i(t)} & i>1 \\
   \end{cases} 
   \right\} \ ,
\end{equation}
where image A is the reference. If image B, C or D is the reference, then the factor $A$ is multiplied by $10^{-0.4 P_i(t)}$ and the reference curve has no magnification. 

The best fit values for the point mass lens are $t_0 =54253.5^{+1.5}_{-1.5}$, $t_E =106.9^{+3.4}_{-3.2}$ days and $u_0 =1.242^{+0.017}_{-0.016}$. These values, particularly the crossing time, $t_E$, can be used to estimate the mass of the lens that caused this event. To do this requires the relative velocity of the object to the source and the observer, as well as the distances to the source and lens. Assuming $\Lambda$CDM cosmology with $H_0 = 67.8 \pm 0.9$ kms\textsuperscript{-1}Mpc\textsuperscript{-1} and $\Omega_m = 0.308 \pm 0.012 $ \citep[][]{Planck2016}, the distance to the lens is $D_L = 2600.9$ Mpc for $z_L=0.454$ \citep[][]{Wisotzki2002} and the distance to the source is $D_S = 12.99$ Gpc for $z_S=2.375$.

\begin{landscape}

\begin{table}

  \caption{COSMOGRAIL Quasars}
    \label{tab:cosmograil}

    \def\arraystretch{1.5}
    \setlength{\tabcolsep}{4pt}
    \begin{threeparttable}
    
  \begin{tabular}{ccccccccccccc}

  \hline
\multicolumn{1}{c}{Object} & \multicolumn{1}{c}{Images}& \multicolumn{2}{c}{Time Delays (Days)}& \multicolumn{7}{c}{Microlensing Coeff. (Mag)} & \multicolumn{1}{c}{Ref.} \\
\cmidrule(r){3-4}
\cmidrule(r){5-11}
  & & {This Work}  & {COSMOGRAIL}  &  {$P_{i,0} $}&  {$P_{i,1} $}&  {$P_{i,2} $}&  {$P_{i,3} $}&  {$P_{i,4} $}&  {$P_{i,5} $}&  {$P_{i,6} $}\\
    \hline

 2M 1134-2103 & AB & $-29.53^{+0.63}_{-0.59}$ & $-30.5 \pm 2.2$	  & $0.03835^{+0.00047}_{-0.00044}$ & $-0.0027^{+0.0013}_{-0.0012}$ & $-0.0107^{+0.0035}_{-0.0035}$ & $0.0259^{+0.0030}_{-0.0030}$ &  $-0.0009^{+0.0046}_{-0.0049}$&- &- &\citet{Millon2020b} \\
 & AC & $8.78^{+0.56}_{-0.56}$ & $8.6 \pm 1.4$ & $0.06502^{+0.00039}_{-0.00039}$ & $0.0021^{+0.0011}_{-0.0011}$ & $-0.0015^{+0.0027}_{-0.0027}$ & $0.0003^{+0.0018}_{-0.0017}$ & $0.0033^{+0.0031}_{-0.0032}$ & -& -&  \\
 & AD & $-65.6^{+2.6}_{-2.3}$ & $-71.9 \pm 6.7 $ & $1.7559^{+0.0011}_{-0.0011}$ & $-0.0110^{+0.0064}_{-0.0058}$ & $0.0151^{+0.0083}_{-0.0084}$ & $0.038^{+0.015}_{-0.016}$ & $-0.028^{+0.015}_{-0.014}$ & -&- & \\
 & BC & $38.60^{+0.49}_{-0.50}$ & $38.9 \pm 2.2$ & $0.02791^{+0.00043}_{-0.00042}$ & $0.0005^{+0.0019}_{-0.0019}$ & $-0.0202^{+0.0030}_{-0.0030}$ & $-0.0156^{+0.0056}_{-0.0056}$ & $0.0137^{+0.0056}_{-0.0058}$ &- &- &  \\
 & BD & $-36.4^{+1.9}_{-2.0}$ & $	-41.5 \pm 7.6$ & $1.7186^{+0.0011}_{-0.0011}$ & $-0.0288^{+0.0065}_{-0.0066}$ & $0.0592^{+0.0077}_{-0.0078}$ & $0.028^{+0.014}_{-0.013}$ & $-0.048^{+0.013}_{-0.013}$ &- &- & \\
 & CD & $-75.6^{+2.4}_{-2.2}$ & $	-80.5 \pm 6.8$ & $1.6927^{+0.0011}_{-0.0011}$ & $-0.0153^{+0.0057}_{-0.0056}$ & $0.0092^{+0.0096}_{-0.0096}$ & $0.059^{+0.016}_{-0.018}$ & $-0.040^{+0.015}_{-0.015}$& -&- &  \\ 
    \hline
DES J0407-5006 & AB & $-123.9^{+1.5}_{-1.5}$ & $ -128.4 \pm 3.6$ & $1.2479^{+0.0016}_{-0.0016}$ & $-0.070^{+0.014}_{-0.013}$ & $0.126^{+0.023}_{-0.024}$ & $-0.010^{+0.012}_{-0.012}$ & $-0.051^{+0.016}_{-0.016}$&- &-  & \citet{Millon2020b} \\
    \hline
DES J0408-5354 & AB & $-113.77^{+0.79}_{-0.76}$ & $-112.1 \pm 2.1$ & $-0.6047^{+0.0054}_{-0.0059}$ & $-0.012^{+0.036}_{-0.036}$ & $-0.072^{+0.054}_{-0.052}$ & $0.008^{+0.042}_{-0.041}$ & $0.042^{+0.041}_{-0.048}$ &- &- & \citet{Courbin2018} \\
 & AD & $-159.1^{+1.8}_{-1.7}$ & $-155.5 \pm 12.8$ & $0.511^{+0.020}_{-0.021}$ & $-0.017^{+0.065}_{-0.066}$ & $-0.134^{+0.072}_{-0.068}$ & $-0.017^{+0.094}_{-0.010}$ & $0.186^{+0.074}_{-0.067}$ &- &- &  \\
 & BD & $-47.6^{+1.3}_{-1.4}$ & $-42.4 \pm 17.6$ & $1.1244^{+0.0043}_{-0.0039}$ & $-0.0643^{+0.0092}_{-0.0087}$ &$-0.096^{+0.027}_{-0.029}$ & $0.111^{+0.023}_{-0.022}$ & $0.077^{+0.031}_{-0.031}$ &- &- \\
     \hline
DES 2325-5229 & AB & $44.83^{+0.90}_{-0.87}$ & $ 43.9 \pm 4.2$ & $1.0985^{+0.0040}_{-0.0041}$ & $-0.131^{+0.011}_{-0.011}$ & $0.045^{+0.037}_{-0.037}$ & $0.397^{+0.033}_{-0.032}$ & $0.037^{+0.050}_{-0.049}$ &- &-  &  \citet{Millon2020b} \\
     \hline
 HE 0047-1756 & AB & $-11.00^{+0.39}_{-0.39}$ & $ -10.8 \pm 1.0$ & $1.6127^{+0.0014}_{-0.0013}$ & $-0.0140^{+0.0029}_{-0.0029}$ & $0.0617^{+0.0060}_{-0.0064}$ & $0.0235^{+0.0040}_{-0.0040}$ & $-0.0550^{+0.0066}_{-0.0063}$ &- &-  &  \citet{Millon2020b} \\
     \hline
 HE 0230-2130 & AC & $17.7^{+2.1}_{-2.1}$ & $ 15.7 \pm 3.9$ & $1.5081^{+0.0070}_{-0.0068}$ & $-0.129^{+0.013}_{-0.013}$ & $0.202^{+0.047}_{-0.047}$ & $-0.002^{+0.017}_{-0.018}$ & $-0.139^{+0.049}_{-0.051}$  &- &-  &  \citet{Millon2020a} \\ 
 & AD & $-10.5^{+9.8}_{-6.1}$ & - & $2.321^{+0.014}_{-0.016}$ & $-0.107^{+0.024}_{-0.024}$ &  $0.183^{+0.089}_{-0.086}$ &$-0.098^{+0.032}_{-0.031}$ & $-0.226^{+0.089}_{-0.092}$ &- &- & \\
 & CD & $-26.9^{+11.4}_{-7.1}$ &  -& $0.802^{+0.014}_{-0.015}$ & $0.035^{+0.029}_{-0.029}$ & $0.004^{+0.010}_{-0.093}$ & $-0.111^{+0.039}_{-0.039}$ & $-0.11^{+0.10}_{-0.10}$ & -&- & \\ 

    \hline
HE 0435-1223 & AB & $-8.00^{+0.37}_{-0.36}$ & $-9.0 \pm 0.8$ & $0.2555^{+0.0019}_{-0.0019}$ & $-0.4259^{+0.0073}_{-0.0068}$ & $0.887^{+0.021}_{-0.021}$ & $0.793^{+0.024}_{-0.025}$ & $-1.910^{+0.056}_{-0.055}$ &  $0.546^{+0.020}_{-0.020}$ &  $0.932^{+0.038}_{-0.039}$  & \citet{Millon2020a} \\
 & AC & $-0.12^{+0.26}_{-0.20}$ & $-0.8 \pm 0.7$ & $0.6614^{+0.0017}_{-0.0017}$ & $-0.4137^{+0.0068}_{-0.0068}$ & $0.773^{+0.020}_{-0.019}$ & $0.497^{+0.023}_{-0.023}$ & $-1.438^{+0.049}_{-0.052}$  & $-0.218^{+0.018}_{-0.019}$ & $0.674^{+0.036}_{-0.033}$ & \\
 & AD & $-13.99^{+0.19}_{-0.27}$ & $-13.8 \pm 0.8 $ & $1.2895^{+0.0019}_{-0.0020}$ & $-0.1620^{+0.0074}_{-0.0074}$ & $0.781^{+0.022}_{-0.022}$ & $0.369^{+0.026}_{-0.025}$ & $-1.678^{+0.059}_{-0.060}$  & $-0.202^{+0.020}_{-0.021}$ & $0.902^{+0.041}_{-0.041}$ & \\
 & BC & $9.06^{+0.20}_{-0.28}$ & $7.8 \pm 0.9$ & $0.4069^{+0.0018}_{-0.0018}$ & $0.0076^{+0.0051}_{-0.0054}$ & $-0.121^{+0.019}_{-0.019}$ & $-0.275^{+0.019}_{-0.018}$ & $0.480^{+0.049}_{-0.049}$ & $0.314^{+0.016}_{-0.016}$ & $-0.257^{+0.033}_{-0.034}$ &\\
 & BD & $-5.22^{+0.39}_{-0.41}$ & $-5.4 \pm 0.8$ & $1.0326^{+0.0020}_{-0.0020}$ & $0.2693^{+0.0059}_{-0.0061}$ & $-0.101^{+0.022}_{-0.023}$ & $-0.448^{+0.021}_{-0.021}$ & $0.228^{+0.058}_{-0.056}$ & $0.363^{+0.018}_{-0.018}$ & $-0.032^{+0.039}_{-0.040}$   \\
 & CD & $-13.99^{+0.19}_{-0.30}$ & $-13.2 \pm 0.8$ & $0.6254^{+0.0019}_{-0.0019}$ & $0.2609^{+0.0057}_{-0.0057}$ & $0.017^{+0.020}_{-0.020}$ & $-0.163^{+0.020}_{-0.021}$ & $-0.247^{+0.050}_{-0.051}$ & $0.041^{+0.018}_{-0.017}$ & $0.229^{+0.036}_{-0.035}$  \\    
    \hline
HE 2149-2745  & AB & $-71.3^{+3.3}_{-3.4}$ & $-39.0 \pm 15.8$	& $1.4350^{+0.0018}_{-0.0018}$ & $-0.0390^{+0.0050}_{-0.0050}$ & $-0.184^{+0.014}_{-0.013}$& $-0.1197^{+0.0078}_{-0.0081}$& $0.175^{+0.015}_{-0.016}$ &- &-   & \citet{Millon2020a} \\
      \hline
HS 0818+1227 & AB & $-39.8^{+3.3}_{-3.2}$ & $-153.8 \pm 13.9$	 & $1.9478^{+0.0040}_{-0.0041}$ & $0.1287^{+0.0081}_{-0.0080}$ & $0.102^{+0.023}_{-0.023}$ & $-0.038^{+0.011}_{-0.011}$ & $-0.116^{+0.023}_{-0.024}$  &- &-& \citet{Millon2020a} \\
      \hline
HS 2209+1914  & AB & $24.0^{+2.4}_{-2.4} $ & $20.0 \pm 5.0$	 & $0.21699^{+0.000053}_{-0.00055}$ & $0.0196^{+0.0030}_{-0.0029}$ & $-0.0071^{+0.0064}_{-0.0066}$ & $0.0870^{+0.0059}_{-0.0059}$ & $0.0707^{+0.0092}_{-0.0091}$ &- &- & \citet{Eulaers2013} \\
    \hline
Q J0158-4325  & AB & $-34.9^{+3.2}_{-3.0} $ & $-22.7 \pm 3.6	$	 &  $1.1236^{+0.0031}_{-0.0031}$ & $0.6289^{+0.0071}_{-0.0072}$ &  $-0.148^{+0.020}_{-0.020}$ & $-0.075^{+0.011}_{-0.010}$ & $-0.027^{+0.022}_{-0.022}$ &- &- & \citet{Millon2020a} \\
    \hline
SDSS J0246-0825  & AB & $3.1^{+2.0}_{-2.0} $ & $0.8 \pm 5.1$	 & $1.8400^{+0.0051}_{-0.0050}$ & $0.476^{+0.013}_{-0.013}$& $-0.892^{+0.035}_{-0.034}$ & $-0.314^{+0.020}_{-0.020}$ & $0.673^{+0.037}_{-0.038}$ &- &- & \citet{Millon2020a} \\
    \hline
SDSS J0832+0404  & AB & $-136.5^{+1.8}_{-1.9}$ & $-125.3 \pm 18.1$	 & $0.9912^{+0.0064}_{-0.0066}$ &$-0.066^{+0.018}_{-0.019}$& $-0.009^{+0.042}_{-0.043}$ & $-0.416^{+0.026}_{-0.024}$ & $0.346^{+0.048}_{-0.047}$ &- &- & \citet{Millon2020a} \\
    \hline
SDSS J0924+0219  & AB & $0.6^{+2.7}_{-2.6}$ & $2.4 \pm 3.8$	 & $1.3827^{+0.0055}_{-0.0056}$ & $-0.014^{+0.015}_{-0.015}$ & $0.291^{+0.038}_{-0.039}$ & $0.476^{+0.023}_{-0.024}$ & $-0.693^{+0.046}_{-0.047}$ &- &- & \citet{Millon2020a} \\
 & AC & $-30.9^{+7.9}_{-8.0}$ & - & $2.785^{+0.017}_{-0.017}$ & $0.082^{+0.047}_{-0.045}$ & $0.14^{+0.11}_{-0.11}$ & $0.827^{+0.081}_{-0.077}$ & $-0.54^{+0.14}_{-0.13}$ &- &- & \\
 & BC & $-32.5^{+8.4}_{-8.5}$ & - & $1.401^{+0.018}_{-0.018}$ & $0.099^{+0.049}_{-0.046}$ & $-0.13^{+0.12}_{-0.12}$ & $0.325^{+0.081}_{-0.083}$ & $0.16^{+0.15}_{-0.14}$  &- &- &  \\

    \hline
  \end{tabular}

  \end{threeparttable}
 \end{table}
\end{landscape}

\begin{landscape}

\begin{table}
\ContinuedFloat  

  \caption{COSMOGRAIL Quasars continued}

    \def\arraystretch{1.5}
    \setlength{\tabcolsep}{4pt}
    \begin{threeparttable}
    
  \begin{tabular}{ccccccccccccc}

  \hline
\multicolumn{1}{c}{Object} & \multicolumn{1}{c}{Images}& \multicolumn{2}{c}{Time Delays (Days)}& \multicolumn{7}{c}{Microlensing Coeff. (Mag)} & \multicolumn{1}{c}{Ref.} \\
\cmidrule(r){3-4}
\cmidrule(r){5-11}
  & & {This Work}  & {COSMOGRAIL}  &  {$P_{i,0} $}&  {$P_{i,1} $}&  {$P_{i,2} $}&  {$P_{i,3} $}&  {$P_{i,4} $}&  {$P_{i,5} $}&  {$P_{i,6} $}\\
      \hline
SDSS J1001+5027 & AB & $-119.9^{+1.4}_{-1.5}$ & $-119.3 \pm 3.3$ & $0.42039^{+0.00010}_{-0.00093}$ & $0.0004^{+0.0039}_{-0.00041}$ & $0.0069^{+0.0089}_{-0.00091}$ & $-0.0215^{+0.0081}_{-0.0080}$& $-0.024^{+0.014}_{-0.013}$ &- &-  & \citet{RathnaKumar2013} \\
     \hline
SDSS J1206+4332 & AB & $-113.0^{+1.1}_{-1.2}$ & $-111.3 \pm 3.0$ & $0.3332^{+0.0023}_{-0.0022}$ & $-0.1670^{+0.0075}_{-0.0073}$ & $-0.213^{+0.016}_{-0.016}$ & $0.104^{+0.015}_{-0.015}$ & $0.141^{+0.023}_{-0.023}$&- &- & \citet{Eulaers2013} \\
    \hline
SDSS J1226-0006 & AB & $25.63^{+0.89}_{-0.76}$ & -	 & $0.9113^{+0.0025}_{-0.0024}$& $0.2639^{+0.0055}_{-0.0056}$ & $0.160^{+0.016}_{-0.016}$  & $-0.1019^{+0.0079}_{-0.0080}$& $-0.082^{+0.017}_{-0.017}$&- &- & \citet{Millon2020a} \\
     \hline
SDSS J1320+1644& AB & $-157.1^{+14.2}_{-14.9}$ & -	 & $0.894^{+0.031}_{-0.028}$ & $0.178^{+0.053}_{-0.066}$  & $-0.24^{+0.24}_{-0.39}$  & $-0.025^{+0.066}_{-0.070}$   & $0.04^{+0.28}_{-0.19}$ &- &-& \citet{Millon2020a} \\
      \hline
SDSS J1322+1052 & AB & $105.6^{+10.0}_{-11.5}$ & -	& $1.6613^{+0.0068}_{-0.0070}$ & $-0.077^{+0.023}_{-0.024}$ & $-0.474^{+0.053}_{-0.053}$ & $0.223^{+0.040}_{-0.039}$ & $0.581^{+0.063}_{-0.065}$&- &- & \citet{Millon2020a} \\
     \hline
SDSS J1335+0118 & AB & $-59.0^{+4.1}_{-4.2}$ & $-56.0 \pm 5.9$ & $1.0958^{+0.0020}_{-0.0021}$ & $-0.1001^{+0.0046}_{-0.0046}$ & $-0.038^{+0.013}_{-0.013}$ & $0.0144^{+0.0065}_{-0.0072}$ & $0.051^{+0.014}_{-0.014}$&- &- & \citet{Millon2020a} \\
     \hline
SDSS J1349+1227 & AB & $-140.8^{+7.5}_{-6.5}$ & - & $1.3180^{+0.0056}_{-0.0069}$ & $-0.007^{+0.013}_{-0.012}$ & $0.020^{+0.039}_{-0.036}$ & $0.089^{+0.018}_{-0.018}$ & $0.005^{+0.039}_{-0.041}$ &- &-& \citet{Millon2020a} \\
     \hline
SDSS J1405+0959 & AB & $-82.8^{+12.3}_{-19.7}$ & -& $0.778^{+0.011}_{-0.010}$ & $0.019^{+0.024}_{-0.024}$ & $0.023^{+0.060}_{-0.061}$  & $-0.004^{+0.033}_{-0.034}$ & $0.035^{+0.069}_{-0.069}$ &- &- & \citet{Millon2020a} \\
     \hline
SDSS J1455+1447 & AB & $-43.4^{+2.3}_{-2.2}$ & $-47.2 \pm 7.6$  & $1.0711^{+0.0041}_{-0.0039}$ & $0.0426^{+0.0084}_{-0.0088}$ & $0.002^{+0.026}_{-0.026}$ & $0.007^{+0.012}_{-0.012}$ & $0.024^{+0.027}_{-0.028}$&- &- & \citet{Millon2020a} \\
     \hline
SDSS J1515+1511 & AB & $-208.3^{+1.8}_{-1.8}$ & $-210.2 \pm 5.6$ & $0.3679^{+0.0022}_{-0.0024}$ & $0.0117^{+0.0070}_{-0.0067}$ & $-0.027^{+0.021}_{-0.019}$ & $-0.025^{+0.019}_{-0.019}$ & $0.053^{+0.032}_{-0.035}$&- &- & \citet{Millon2020a} \\
\hline

SDSS J1620+1203 & AB & $-158.3^{+2.4}_{-2.5}$ & $-171.5 \pm 8.7$& $1.2387^{+0.0064}_{-0.0067}$ & $0.003^{+0.016}_{-0.016}$ & $-0.082^{+0.048}_{-0.045}$ & $-0.093^{+0.037}_{-0.036}$ & $0.148^{+0.061}_{-0.062}$&- &- & \citet{Millon2020a} \\
     \hline
PG 1115+080 & AB & $-6.90^{+0.68}_{-0.69}$ & $-8.3 \pm 1.5$ & $2.5709^{+0.0018}_{-0.0018}$ & $0.0211^{+0.0041}_{-0.0041}$ & $0.037^{+0.012}_{-0.012}$ & $-0.0103^{+0.0060}_{-0.0059}$ & $-0.034^{+0.013}_{-0.012}$&- &- & \citet{Bonvin2018} \\
 & AC  & $9.43^{+0.83}_{-0.82}$ & $9.9 \pm 1.1$ & $2.2193^{+0.0017}_{-0.0016}$ & $0.0040^{+0.0035}_{-0.0035}$ & $0.033^{+0.010}_{-0.010}$ & $-0.0048^{+0.0066}_{-0.0063}$ & $-0.038^{+0.012}_{-0.012}$&- &- &   \\
 & BC & $-16.35^{+0.94}_{-0.90}$ & $18.8 \pm 1.6$ & $-0.3569^{+0.0022}_{-0.0023}$ & $-0.0268^{+0.0053}_{-0.0051}$ & $0.012^{+0.012}_{-0.011}$ & $0.0255^{+0.0098}_{-0.0099}$ & $-0.009^{+0.014}_{-0.014}$&- &- &  \\ 
    \hline
PSJ 1606-2333& AB & $-12.9^{+1.1}_{-1.1}$ & $	-10.4 \pm  2.2$	 & $0.1523^{+0.0029}_{-0.0030}$ & $0.0342^{+0.0046}_{-0.0048}$ & $0.001^{+0.012}_{-0.011}$ & $-0.0186^{+0.0074}_{-0.0074}$ & $-0.011^{+0.012}_{-0.013}$ &- &-& \citet{Millon2020b} \\
 & AC & $-30.6^{+2.2}_{-2.4}$ & $-29.2 \pm  4.7$ & $0.5593^{+0.0059}_{-0.0060}$ & $0.0338^{+0.0050}_{-0.0053}$ & $-0.056^{+0.022}_{-0.023}$ & $0.001^{+0.010}_{-0.010}$ & $0.003^{+0.023}_{-0.023}$ &- &-&  \\
 & AD & $-40.1^{+2.8}_{-2.4}$ & $-45.7 \pm  10.9$ & $0.7449^{+0.0071}_{-0.0069}$ & $0.0689^{+0.0066}_{-0.0064}$ & $-0.037^{+0.025}_{-0.025}$ & $-0.004^{+0.014}_{-0.014}$  & $-0.022^{+0.024}_{-0.024}$ &- &-&  \\
 & BC & $-17.5^{+1.8}_{-2.0}$ & $	-19.3 \pm  4.5$ & $0.4114^{+0.0045}_{-0.0049}$ & $-0.0011^{+0.0054}_{-0.0054}$ & $-0.056^{+0.021}_{-0.020}$ & $0.0161^{+0.0090}_{-0.0091}$ & $0.005^{+0.020}_{-0.021}$  &- &-&  \\
 & BD & $-27.4^{+3.2}_{-2.7}$ & $-36.2 \pm  10.4$ & $0.5986^{+0.0064}_{-0.0061}$ & $0.0349^{+0.0064}_{-0.0066}$ & $-0.038^{+0.024}_{-0.027}$ & $0.0119^{+0.013}_{-0.012}$ & $-0.023^{+0.025}_{-0.024}$ &- &-&   \\
 & CD & $-9.0^{+3.9}_{-3.4}$ & $	-13.8 \pm  8.4$ & $0.1874^{+0.0084}_{-0.0073}$ & $0.0262^{+0.0066}_{-0.0075}$ & $0.034^{+0.028}_{-0.032}$ & $0.002^{+0.011}_{-0.010}$ & $-0.047^{+0.026}_{-0.025}$  &- &-&  \\ 
    \hline
Q 1355-2257 & AB & $-64.5^{+4.6}_{-4.7}$ & $-81.5 \pm 11.4$ & $1.5528^{+0.0049}_{-0.0049}$ & $-0.120^{+0.015}_{-0.015}$ & $0.287^{+0.039}_{-0.039}$ & $0.005^{+0.037}_{-0.037}$ & $-0.259^{+0.059}_{-0.058}$ &- &-& \citet{Millon2020a} \\
    \hline
Q 2237+0305 & AB & $11.7^{+4.7}_{-5.4}$ & - & $0.8885^{+0.0050}_{-0.0051}$ & $-0.128^{+0.016}_{-0.016}$ & $0.139^{+0.034}_{-0.034}$ & $0.021^{+0.023}_{-0.022}$ & $-0.127^{+0.037}_{-0.035}$ &- &- & \citet{Millon2020a} \\
  \hline
UM 673 & AB & $7.7^{+10.6}_{-13.0}$ & $-97.7 \pm 15.8$	& $2.1667^{+0.0022}_{-0.0022}$ & $0.0170^{+0.0066}_{-0.0066}$ & $-0.127^{+0.015}_{-0.015}$ & $-0.008^{+0.010}_{-0.010}$ & $0.176^{+0.016}_{-0.016}$ &- &- &  \citet{Millon2020a} \\

  \hline
  \end{tabular}

  \end{threeparttable}
 \end{table}
\end{landscape}

\begin{landscape}

\begin{table}
\ContinuedFloat  

  \caption{COSMOGRAIL Quasars continued}

    \def\arraystretch{1.5}
    \setlength{\tabcolsep}{4pt}
    \begin{threeparttable}
    
  \begin{tabular}{ccccccccccccc}

  \hline
\multicolumn{1}{c}{Object} & \multicolumn{1}{c}{Images}& \multicolumn{2}{c}{Time Delays (Days)}& \multicolumn{7}{c}{Microlensing Coeff. (Mag)} & \multicolumn{1}{c}{Ref.} \\
\cmidrule(r){3-4}
\cmidrule(r){5-11}
  & & {This Work}  & {COSMOGRAIL}  &  {$P_{i,0} $}&  {$P_{i,1} $}&  {$P_{i,2} $}&  {$P_{i,3} $}&  {$P_{i,4} $}&  {$P_{i,5} $}&  {$P_{i,6} $}\\
  \hline
WFI J2026-4536 & AB & $4.4^{+2.1}_{-2.3}$ & $18.0 \pm 4.8$	& $1.9398^{+0.0027}_{-0.0029}$ & $-0.1713^{+0.0075}_{-0.0075}$ & $0.255^{+0.018}_{-0.019}$ & $0.108^{+0.011}_{-0.012}$ & $-0.224^{+0.021}_{-0.021}$&- &- &  \citet{Millon2020a} \\
 & AC & $2.6^{+4.2}_{-4.6}$ & - & $2.0434^{+0.0050}_{-0.0050}$ & $-0.031^{+0.014}_{-0.014}$ & $0.246^{+0.035}_{-0.036}$ & $-0.023^{+0.022}_{-0.022}$ & $-0.196^{+0.041}_{-0.041}$ &- &-& \\
 & BC & $2.2^{+5.1}_{-5.5}$ & - & $0.1031^{+0.0056}_{-0.0055}$ & $0.144^{+0.015}_{-0.016}$ & $-0.006^{+0.039}_{-0.039}$ & $0.137^{+0.025}_{-0.024}$ & $0.025^{+0.046}_{-0.045}$ &- &-& \\
  \hline
WFI 2033-4723 & AB & $40.54^{+0.60}_{-0.64}$ & $36.2 \pm 0.7$	& $1.0681^{+0.0013}_{-0.0012}$ & $-0.0072^{+0.0030}_{-0.0030}$ & $0.0349^{+0.0083}_{-0.0081}$ & $0.0390^{+0.0052}_{-0.0050}$ & $0.019^{+0.010}_{-0.010}$ &- &-&  \citet{Bonvin2019} \\
 & AC & $-27.23^{+0.69}_{-0.66}$ & $	-23.3 \pm 1.3$ & $1.3830^{+0.0014}_{-0.0014}$ & $0.0766^{+0.0033}_{-0.0033}$ & $-0.0752^{+0.0086}_{-0.0091}$ & $-0.0077^{+0.0054}_{-0.0055}$ & $0.000^{+0.010}_{-0.011}$&- &- & \\
 & BC & $-68.3^{+1.2}_{-0.8}$ & $	-59.4 \pm 1.3$ & $0.3146^{+0.0016}_{-0.0016}$ & $0.0859^{+0.0041}_{-0.0040}$ & $-0.106^{+0.011}_{-0.012}$ & $-0.0456^{+0.0066}_{-0.0069}$ & $-0.021^{+0.014}_{-0.014}$&- &- &\\ 
  \hline
 WG 0214-2105 & AB & $10.24^{+0.67}_{-0.64}$ & $7.5 \pm 2.8$ & $-0.0773^{+0.0043}_{-0.0042}$ & $-0.1127^{+0.0052}_{-0.0054}$ & $0.031^{+0.019}_{-0.019}$ & $0.0632^{+0.0080}_{-0.0076}$ & $-0.0635^{+0.0018}_{-0.0018}$ &- &- & \citet{Millon2020b} \\
 & AC & $-5.82^{+0.69}_{-0.69}$ & $-6.7 \pm 3.6$ & $-0.0413^{+0.0041}_{-0.0041}$ & $-0.0320^{+0.0057}_{-0.0057}$ & $-0.020^{+0.019}_{-0.019}$ & $0.0118^{+0.0083}_{-0.0086}$ & $0.012^{+0.018}_{-0.018}$  &- &-&  \\
 & AD & $-43.2^{+1.8}_{-1.8}$ & $-14.1 \pm 4.9$ & $0.6228^{+0.0079}_{-0.0078}$ & $-0.111^{+0.013}_{-0.013}$ & $0.403^{+0.036}_{-0.035}$ & $0.241^{+0.023}_{-0.023}$ & $-0.339^{+0.038}_{-0.039}$ &- &-&   \\
 & BC & $-16.01^{+0.56}_{-0.59}$ & $-14.2 \pm 2.6$ & $0.0314^{+0.0043}_{-0.0043}$ & $0.0840^{+0.0058}_{-0.0058}$ & $-0.047^{+0.019}_{-0.019}$ & $-0.0657^{+0.0093}_{-0.0090}$ & $0.083^{+0.020}_{-0.019}$ &- &-&   \\
 & BD & $-53.5^{+1.7}_{-1.7}$ & $-21.6 \pm 4.6$ & $0.6812^{+0.0079}_{-0.0078}$ & $0.007^{+0.014}_{-0.014}$ & $0.413^{+0.036}_{-0.037}$ & $0.131^{+0.024}_{-0.024}$ & $-0.271^{+0.040}_{-0.041}$ &- &-&    \\
 & CD & $-37.4^{+1.6}_{-1.6}$ & $-7.5 \pm 4.7$ & $0.6616^{+0.0074}_{-0.0077}$ & $-0.083^{+0.012}_{-0.012}$ & $0.422^{+0.035}_{-0.035}$ & $0.232^{+0.020}_{-0.020}$ & $-0.346^{+0.037}_{-0.038}$ &- &-&   \\ 
  \hline
  \end{tabular}

  \end{threeparttable}
 \end{table}
\end{landscape}


From the crossing time the Einstein ring radius can be estimated, which relates to the mass of the lens, $M_L$ from the following

\begin{equation}
      \frac{t_E v_{\textrm{rel}}}{(1+z_L)D_L} \approx \theta_E = \sqrt{\frac{4 G M_L}{c^2} \frac{D_S - D_L}{D_S D_L}},
\end{equation}
where the factor of $1+z_L$ accounts for cosmic time dilation on the measured crossing time. To estimate $v_{\textrm{rel}}$, we would require the velocity of the source, lens and observer however we can construct a prior on $v_{\textrm{rel}}$ making some approximations, following \citet{Poindexter2010} and \citet{Blackburne2014}. Namely we assume that the source velocity is negligible compared to the lens velocity due to cosmic time dilation/geometric projection effects and then construct a Gaussian prior on the relative velocity measured as the lens velocity relative to the observers velocity, estimated from the projection of the cosmic microwave background (CMB) dipole \citep{Hinshaw2009}. The width of this prior is estimated from the sum of the velocity dispersion's of the source and lens galaxies to account for the random motion of stars which cause these microlensing events. For HE 0435-1223 this has been estimated to be $\sigma_S = 227$ km s\textsuperscript{-1} and $\sigma_L = 277$ km s\textsuperscript{-1} for the source and the lens respectively \citep{Blackburne2014}. Therefore the prior for the relative velocity is 
\begin{equation}
    P(\pmb{v}_{\textrm{rel}}) \propto \exp \left(-\frac{\pmb{v}_{\textrm{rel}} - \pmb{v}_{\textrm{CMB}}}{2\sigma^2}\right),
\end{equation}
where $\pmb{v}_{\textrm{CMB}}$ is the CMB dipole velocity projected onto the lens plane, which for HE 0435-1223 is (363, -56) km s\textsuperscript{-1} east and north \citep[][]{Blackburne2014} and the velocity dispersion is given by
\begin{equation}
    \sigma^2 = \sigma_L^2 + \left( \sigma_S \frac{1+z_L}{1+z_S} \frac{D_L}{D_S}\right)^2 = (290 \ \textrm{km s\textsuperscript{-1}})^2 .
\end{equation}

To achieve a probability distribution of the mass of the lens, we generated a 2d probability distribution for $\pmb{v}_{\textrm{rel}}$ by sampling Gaussian random numbers with a mean of the (CMB) dipole velocity for each component and a standard deviation of 290 km s\textsuperscript{-1}. This is then converted into a distribution on the speed, $v_{\textrm{rel}}$, by taking the magnitude of each of the velocity coordinates and by also sampling Gaussian random numbers for the distribution on $t_E$, we calculated a probability distribution for the lens mass, $M_L$. 

The resulting distribution is highly skewed towards low mass with a median and 16\textsuperscript{th}/84\textsuperscript{th} percentiles giving $M_L = 8^{+11}_{-6} \textrm{M}_{\oplus}$. The low microlensing mass measured  suggests that this object could be a planet - specifically a rogue planet that is not bound to a star. If it did orbit a star, the host star would dominate the microlensing amplification and render the planet undetectable. Such objects have been proposed to explain microlensing activity in RX J1131-1231 \citep{Dai2018} where $\sim$ 2000 objects with masses between the Moon and Jupiter masses were estimated by analysing Fe K$\alpha$ line energy shifts.  Taking a larger estimate of the velocity dispersion of $\sigma = 1000\, \textrm{km s\textsuperscript{-1}}$ results in a larger mass of $M_L = 0.16^{+0.27}_{-0.12}\, M_{\textrm{Jup}}$ which is $\sim$ 6.7 times more massive but still a planet-mass object. The probability of a rogue planet causing this event is extremely low considering their size and suggested abundance compared with stars in the galaxy.

The other possibility is that this event is caused by a star that is moving extremely fast. Taking a typical M-dwarf of $\sim 0.3 \textrm{M}_{\odot}$, would require a speed of $\sim$ 53~000 km s\textsuperscript{-1} which is extremely fast compared to the prior - also an unlikely result. As both of these results are unlikely it suggests that the assumption that this feature in the lightcurve is cause by a point-mass may be inaccurate. The caustic patterns that cause this lensing are likely much more complex than a single point mass and so while it provides a good fit to the lightcurve, any physical interpretation of the parameters are likely inaccurate.

The only object where we were unable to obtain a good fit was for RX J1131-1231, which similar to HE 0435-1223 is a quadruply-imaged quasar \citep{Sluse2003}, that was observed for 15 years. We had difficulty constraining the parameters for this object, specifically with the additional microlensing effect. We tried using a 6\textsuperscript{th} order polynomial as well as including point mass lenses similar to HE 0435-12234, however we were unable to constrain these parameters. This object likely requires a more complex microlensing model. Using a higher order polynomial may work but would become over-flexible at the edges of the data and so a better solution could be using a series of splines that can become more flexible where appropriate to fit certain features of the microlensing lightcurve. This is the approach of \citep[][]{Tewes2013}, however they also use splines to model the quasar variability. We have shown that the running optimal average provides a good model of the intrinsic quasar variability and so modifying our method to use splines for the microlensing variability may be a good option for these objects that have been observed for a very long period of time.

\section{Conclusions} \label{sec:Conclusions}
This paper presents the use of a running optimal average to model the variability of quasars, with the main aim of measuring the time delay between quasar lightcurves. This method can model many lightcurves simultaneously, providing the maximum information for determining the shape of the variability from the running optimal average. We optimise the flexibility of the ROA by calculating the effective number of parameters and minimising the Bayesian Information Criterion (BIC) through MCMC sampling of the joint posterior probability distributions of the parameters. We tested this method with mock data as well as real data in the form of gravitationally lensed quasars that were observed and analysed as part of the COSMOGRAIL project. From this testing the main findings are:

\begin{enumerate}
    \item Fitting to mock data with high S/N, low S/N and large gaps, PyROA recovers precise time delays, with uncertainties comparable to JAVELIN and significantly smaller than ICCF.
    \item From fitting to 50 mock data sets, generated with different random walks, we verified that the uncertainties on the time delay parameters were accurate, with the normalised residuals forming a normal distribution with a mean of zero and rms of one.  
    \item Our method is easily able to deal with large gaps in individual lightcurves, either where data points from another lightcurve provides information within the gap of another, or the ROA interpolates across the gap with an error envelope that expands accordingly.
    \item By including a noise model that allows the variance of flux measurements to increase, PyROA is able to recover accurate time delays when the flux errors are deliberately underestimated, while JAVELIN fails.
    \item By including microlensing effects, we modelled the lightcurves of 33 gravitationally lensed quasars from the COSMOGRAIL project. We find delays that are consistent with the previous analysis, with the exception of a few objects. We consistently finds smaller errors for the time delays between images as well as find delays for data that previously were unable to yield a measurement.

\end{enumerate}

In addition to measuring the time delays between lightcurves, PyROA provides a model of the driving lightcurve which can be used for further analysis. For example this can be used to generate a power-density spectrum of the lightcurves, decompose lightcurves into variable/fixed components to separate AGN from galaxy or inter-calibrate data from multiple telescopes where the ROA provides a model of the merged lightcurve. The code PyROA is publicly available, providing a new tool in reverberation mapping studies and gravitationally lensed quasars.

\section*{Acknowledgements}
KH and JVHS acknowledge support from STFC grant ST/R000824/1.

\section*{Data Availability}

The lightcurves for the COSMOGRAIL quasars were obtained from \url{https://obswww.unige.ch/~millon/d3cs/COSMOGRAIL_public/}. Our results for each object including a plot of the model/data, a corner plot, the MCMC samples and the individual lightcurves used for the fit can be found at \url{https://dx.doi.org/10.5281/zenodo.5060008}.



\bibliographystyle{mnras}
\bibliography{References} 

\begin{thebibliography}{}
\makeatletter
\relax
\def\mn@urlcharsother{\let\do\@makeother \do\$\do\&\do\#\do\^\do\_\do\%\do\~}
\def\mn@doi{\begingroup\mn@urlcharsother \@ifnextchar [ {\mn@doi@}
  {\mn@doi@[]}}
\def\mn@doi@[#1]#2{\def\@tempa{#1}\ifx\@tempa\@empty \href
  {http://dx.doi.org/#2} {doi:#2}\else \href {http://dx.doi.org/#2} {#1}\fi
  \endgroup}
\def\mn@eprint#1#2{\mn@eprint@#1:#2::\@nil}
\def\mn@eprint@arXiv#1{\href {http://arxiv.org/abs/#1} {{\tt arXiv:#1}}}
\def\mn@eprint@dblp#1{\href {http://dblp.uni-trier.de/rec/bibtex/#1.xml}
  {dblp:#1}}
\def\mn@eprint@#1:#2:#3:#4\@nil{\def\@tempa {#1}\def\@tempb {#2}\def\@tempc
  {#3}\ifx \@tempc \@empty \let \@tempc \@tempb \let \@tempb \@tempa \fi \ifx
  \@tempb \@empty \def\@tempb {arXiv}\fi \@ifundefined
  {mn@eprint@\@tempb}{\@tempb:\@tempc}{\expandafter \expandafter \csname
  mn@eprint@\@tempb\endcsname \expandafter{\@tempc}}}

\bibitem[\protect\citeauthoryear{{Blackburne}, {Kochanek}, {Chen}, {Dai}  \&
  {Chartas}}{{Blackburne} et~al.}{2014}]{Blackburne2014}
{Blackburne} J.~A.,  {Kochanek} C.~S.,  {Chen} B.,  {Dai} X.,   {Chartas} G.,
  2014, \mn@doi [\apj] {10.1088/0004-637X/789/2/125}, \href
  {https://ui.adsabs.harvard.edu/abs/2014ApJ...789..125B} {789, 125}

\bibitem[\protect\citeauthoryear{{Blandford} \& {McKee}}{{Blandford} \&
  {McKee}}{1982}]{Blandford1982}
{Blandford} R.~D.,  {McKee} C.~F.,  1982, \mn@doi [\apj] {10.1086/159843},
  \href {https://ui.adsabs.harvard.edu/abs/1982ApJ...255..419B} {255, 419}

\bibitem[\protect\citeauthoryear{{Bonvin} et~al.,}{{Bonvin}
  et~al.}{2018}]{Bonvin2018}
{Bonvin} V.,  et~al., 2018, \mn@doi [\aap] {10.1051/0004-6361/201833287}, \href
  {https://ui.adsabs.harvard.edu/abs/2018A&A...616A.183B} {616, A183}

\bibitem[\protect\citeauthoryear{{Bonvin} et~al.,}{{Bonvin}
  et~al.}{2019}]{Bonvin2019}
{Bonvin} V.,  et~al., 2019, \mn@doi [\aap] {10.1051/0004-6361/201935921}, \href
  {https://ui.adsabs.harvard.edu/abs/2019A&A...629A..97B} {629, A97}

\bibitem[\protect\citeauthoryear{Brown et~al.,}{Brown et~al.}{2013}]{Brown2013}
Brown T.~M.,  et~al., 2013, \mn@doi [Publications of the Astronomical Society
  of the Pacific] {10.1086/673168}, 125, 1031

\bibitem[\protect\citeauthoryear{{Cackett}, {Horne}  \& {Winkler}}{{Cackett}
  et~al.}{2007}]{Cackett2007}
{Cackett} E.~M.,  {Horne} K.,   {Winkler} H.,  2007, \mn@doi [\mnras]
  {10.1111/j.1365-2966.2007.12098.x}, \href
  {https://ui.adsabs.harvard.edu/abs/2007MNRAS.380..669C} {380, 669}

\bibitem[\protect\citeauthoryear{{Cackett}, {Bentz}  \& {Kara}}{{Cackett}
  et~al.}{2021}]{Cackett2021}
{Cackett} E.~M.,  {Bentz} M.~C.,   {Kara} E.,  2021, \mn@doi [iScience]
  {10.1016/j.isci.2021.102557}, \href
  {https://ui.adsabs.harvard.edu/abs/2021iSci...24j2557C} {24, 102557}

\bibitem[\protect\citeauthoryear{{Chan}, {Millon}, {Bonvin}  \&
  {Courbin}}{{Chan} et~al.}{2020}]{Chan2020}
{Chan} J.~H.~H.,  {Millon} M.,  {Bonvin} V.,   {Courbin} F.,  2020, \mn@doi
  [\aap] {10.1051/0004-6361/201935423}, \href
  {https://ui.adsabs.harvard.edu/abs/2020A&A...636A..52C} {636, A52}

\bibitem[\protect\citeauthoryear{{Cooke} \& {Kantowski}}{{Cooke} \&
  {Kantowski}}{1975}]{Cooke&Kantowski1975}
{Cooke} J.~H.,  {Kantowski} R.,  1975, \mn@doi [\apjl] {10.1086/181697}, \href
  {https://ui.adsabs.harvard.edu/abs/1975ApJ...195L..11C} {195, L11}

\bibitem[\protect\citeauthoryear{{Courbin} et~al.,}{{Courbin}
  et~al.}{2018}]{Courbin2018}
{Courbin} F.,  et~al., 2018, \mn@doi [\aap] {10.1051/0004-6361/201731461},
  \href {https://ui.adsabs.harvard.edu/abs/2018A&A...609A..71C} {609, A71}

\bibitem[\protect\citeauthoryear{{Dai} \& {Guerras}}{{Dai} \&
  {Guerras}}{2018}]{Dai2018}
{Dai} X.,  {Guerras} E.,  2018, \mn@doi [\apjl] {10.3847/2041-8213/aaa5fb},
  \href {https://ui.adsabs.harvard.edu/abs/2018ApJ...853L..27D} {853, L27}

\bibitem[\protect\citeauthoryear{{Dexter} \& {Agol}}{{Dexter} \&
  {Agol}}{2011}]{Dexter2011}
{Dexter} J.,  {Agol} E.,  2011, \mn@doi [\apjl] {10.1088/2041-8205/727/1/L24},
  \href {https://ui.adsabs.harvard.edu/abs/2011ApJ...727L..24D} {727, L24}

\bibitem[\protect\citeauthoryear{{Edelson} et~al.,}{{Edelson}
  et~al.}{2019}]{Edelson2019}
{Edelson} R.,  et~al., 2019, \mn@doi [\apj] {10.3847/1538-4357/aaf3b4}, \href
  {https://ui.adsabs.harvard.edu/abs/2019ApJ...870..123E} {870, 123}

\bibitem[\protect\citeauthoryear{{Eigenbrod}, {Courbin}, {Vuissoz}, {Meylan},
  {Saha}  \& {Dye}}{{Eigenbrod} et~al.}{2005}]{COSMOGRAIL}
{Eigenbrod} A.,  {Courbin} F.,  {Vuissoz} C.,  {Meylan} G.,  {Saha} P.,   {Dye}
  S.,  2005, \mn@doi [\aap] {10.1051/0004-6361:20042422}, \href
  {https://ui.adsabs.harvard.edu/abs/2005A&A...436...25E} {436, 25}

\bibitem[\protect\citeauthoryear{{Eulaers} et~al.,}{{Eulaers}
  et~al.}{2013}]{Eulaers2013}
{Eulaers} E.,  et~al., 2013, \mn@doi [\aap] {10.1051/0004-6361/201321140},
  \href {https://ui.adsabs.harvard.edu/abs/2013A&A...553A.121E} {553, A121}

\bibitem[\protect\citeauthoryear{{Event Horizon Telescope Collaboration}
  et~al.,}{{Event Horizon Telescope Collaboration} et~al.}{2019}]{EHT2019}
{Event Horizon Telescope Collaboration} et~al., 2019, \mn@doi [\apjl]
  {10.3847/2041-8213/ab0ec7}, \href
  {https://ui.adsabs.harvard.edu/abs/2019ApJ...875L...1E} {875, L1}

\bibitem[\protect\citeauthoryear{{Fabian}}{{Fabian}}{2012}]{Fabian2012}
{Fabian} A.~C.,  2012, \mn@doi [\araa] {10.1146/annurev-astro-081811-125521},
  \href {https://ui.adsabs.harvard.edu/abs/2012ARA&A..50..455F} {50, 455}

\bibitem[\protect\citeauthoryear{{Fausnaugh} et~al.,}{{Fausnaugh}
  et~al.}{2016}]{Fausnaugh2016}
{Fausnaugh} M.~M.,  et~al., 2016, \mn@doi [\apj] {10.3847/0004-637X/821/1/56},
  \href {https://ui.adsabs.harvard.edu/abs/2016ApJ...821...56F} {821, 56}

\bibitem[\protect\citeauthoryear{{Foreman-Mackey}, {Hogg}, {Lang}  \&
  {Goodman}}{{Foreman-Mackey} et~al.}{2013}]{emcee}
{Foreman-Mackey} D.,  {Hogg} D.~W.,  {Lang} D.,   {Goodman} J.,  2013, \mn@doi
  [\pasp] {10.1086/670067}, \href
  {https://ui.adsabs.harvard.edu/abs/2013PASP..125..306F} {125, 306}

\bibitem[\protect\citeauthoryear{{Gaskell} \& {Peterson}}{{Gaskell} \&
  {Peterson}}{1987}]{ICCF}
{Gaskell} C.~M.,  {Peterson} B.~M.,  1987, \mn@doi [\apjs] {10.1086/191216},
  \href {https://ui.adsabs.harvard.edu/abs/1987ApJS...65....1G} {65, 1}

\bibitem[\protect\citeauthoryear{{Gehrels} et~al.,}{{Gehrels}
  et~al.}{2004}]{Gehrels2004}
{Gehrels} N.,  et~al., 2004, \mn@doi [\apj] {10.1086/422091}, \href
  {https://ui.adsabs.harvard.edu/abs/2004ApJ...611.1005G} {611, 1005}

\bibitem[\protect\citeauthoryear{{Gravity Collaboration} et~al.,}{{Gravity
  Collaboration} et~al.}{2018}]{Gravity2018}
{Gravity Collaboration} et~al., 2018, \mn@doi [\nat]
  {10.1038/s41586-018-0731-9}, \href
  {https://ui.adsabs.harvard.edu/abs/2018Natur.563..657G} {563, 657}

\bibitem[\protect\citeauthoryear{{Grier} et~al.,}{{Grier}
  et~al.}{2017}]{Grier2017}
{Grier} C.~J.,  et~al., 2017, \mn@doi [\apj] {10.3847/1538-4357/aa98dc}, \href
  {https://ui.adsabs.harvard.edu/abs/2017ApJ...851...21G} {851, 21}

\bibitem[\protect\citeauthoryear{{Heckman} \& {Best}}{{Heckman} \&
  {Best}}{2014}]{Heckman2014}
{Heckman} T.~M.,  {Best} P.~N.,  2014, \mn@doi [\araa]
  {10.1146/annurev-astro-081913-035722}, \href
  {https://ui.adsabs.harvard.edu/abs/2014ARA&A..52..589H} {52, 589}

\bibitem[\protect\citeauthoryear{{Hern{\'a}ndez Santisteban}
  et~al.,}{{Hern{\'a}ndez Santisteban} et~al.}{2020}]{HernandezSantisteban2020}
{Hern{\'a}ndez Santisteban} J.~V.,  et~al., 2020, \mn@doi [\mnras]
  {10.1093/mnras/staa2365}, \href
  {https://ui.adsabs.harvard.edu/abs/2020MNRAS.498.5399H} {498, 5399}

\bibitem[\protect\citeauthoryear{{Hinshaw} et~al.,}{{Hinshaw}
  et~al.}{2009}]{Hinshaw2009}
{Hinshaw} G.,  et~al., 2009, \mn@doi [\apjs] {10.1088/0067-0049/180/2/225},
  \href {https://ui.adsabs.harvard.edu/abs/2009ApJS..180..225H} {180, 225}

\bibitem[\protect\citeauthoryear{{Homayouni} et~al.,}{{Homayouni}
  et~al.}{2021}]{Homayouni2021}
{Homayouni} Y.,  et~al., 2021, arXiv e-prints, \href
  {https://ui.adsabs.harvard.edu/abs/2021arXiv210502884H} {p. arXiv:2105.02884}

\bibitem[\protect\citeauthoryear{{Horne}, {Peterson}, {Collier}  \&
  {Netzer}}{{Horne} et~al.}{2004}]{Horne2004}
{Horne} K.,  {Peterson} B.~M.,  {Collier} S.~J.,   {Netzer} H.,  2004, \mn@doi
  [\pasp] {10.1086/420755}, \href
  {https://ui.adsabs.harvard.edu/abs/2004PASP..116..465H} {116, 465}

\bibitem[\protect\citeauthoryear{{Kara} et~al.,}{{Kara}
  et~al.}{2021}]{Kara2021}
{Kara} E.,  et~al., 2021, arXiv e-prints, \href
  {https://ui.adsabs.harvard.edu/abs/2021arXiv210505840K} {p. arXiv:2105.05840}

\bibitem[\protect\citeauthoryear{Karson}{Karson}{1968}]{KSTest}
Karson M.,  1968, \mn@doi [Journal of the American Statistical Association]
  {10.1080/01621459.1968.11009335}, 63, 1047

\bibitem[\protect\citeauthoryear{{Kawaguchi}, {Mineshige}, {Umemura}  \&
  {Turner}}{{Kawaguchi} et~al.}{1998}]{Kawaguchi1998}
{Kawaguchi} T.,  {Mineshige} S.,  {Umemura} M.,   {Turner} E.~L.,  1998,
  \mn@doi [\apj] {10.1086/306105}, \href
  {https://ui.adsabs.harvard.edu/abs/1998ApJ...504..671K} {504, 671}

\bibitem[\protect\citeauthoryear{{Koptelova} et~al.,}{{Koptelova}
  et~al.}{2012}]{Koptelova2012}
{Koptelova} E.,  et~al., 2012, \mn@doi [\aap] {10.1051/0004-6361/201116645},
  \href {https://ui.adsabs.harvard.edu/abs/2012A&A...544A..51K} {544, A51}

\bibitem[\protect\citeauthoryear{{Koz{\l}owski} et~al.,}{{Koz{\l}owski}
  et~al.}{2010}]{Kozlowski2010}
{Koz{\l}owski} S.,  et~al., 2010, \mn@doi [\apj] {10.1088/0004-637X/708/2/927},
  \href {https://ui.adsabs.harvard.edu/abs/2010ApJ...708..927K} {708, 927}

\bibitem[\protect\citeauthoryear{{Lin} et~al.,}{{Lin} et~al.}{2017}]{Lin2017}
{Lin} H.,  et~al., 2017, \mn@doi [\apjl] {10.3847/2041-8213/aa624e}, \href
  {https://ui.adsabs.harvard.edu/abs/2017ApJ...838L..15L} {838, L15}

\bibitem[\protect\citeauthoryear{{Lynden-Bell}}{{Lynden-Bell}}{1969}]{LyndenBell1969}
{Lynden-Bell} D.,  1969, \mn@doi [\nat] {10.1038/223690a0}, \href
  {https://ui.adsabs.harvard.edu/abs/1969Natur.223..690L} {223, 690}

\bibitem[\protect\citeauthoryear{{MacLeod} et~al.,}{{MacLeod}
  et~al.}{2010}]{MacLeod2010}
{MacLeod} C.~L.,  et~al., 2010, \mn@doi [\apj] {10.1088/0004-637X/721/2/1014},
  \href {https://ui.adsabs.harvard.edu/abs/2010ApJ...721.1014M} {721, 1014}

\bibitem[\protect\citeauthoryear{{Millon} et~al.,}{{Millon}
  et~al.}{2020a}]{Millon2020a}
{Millon} M.,  et~al., 2020a, \mn@doi [\aap] {10.1051/0004-6361/202037740},
  \href {https://ui.adsabs.harvard.edu/abs/2020A&A...640A.105M} {640, A105}

\bibitem[\protect\citeauthoryear{{Millon} et~al.,}{{Millon}
  et~al.}{2020b}]{Millon2020b}
{Millon} M.,  et~al., 2020b, \mn@doi [\aap] {10.1051/0004-6361/202038698},
  \href {https://ui.adsabs.harvard.edu/abs/2020A&A...642A.193M} {642, A193}

\bibitem[\protect\citeauthoryear{{Oscoz}, {Serra-Ricart}, {Mediavilla}  \&
  {Mu{\~n}oz}}{{Oscoz} et~al.}{2013}]{Oscoz2013}
{Oscoz} A.,  {Serra-Ricart} M.,  {Mediavilla} E.,   {Mu{\~n}oz} J.~A.,  2013,
  \mn@doi [\apj] {10.1088/0004-637X/779/2/144}, \href
  {https://ui.adsabs.harvard.edu/abs/2013ApJ...779..144O} {779, 144}

\bibitem[\protect\citeauthoryear{{Paczynski}}{{Paczynski}}{1986}]{Paczynski1986}
{Paczynski} B.,  1986, \mn@doi [\apj] {10.1086/164140}, \href
  {https://ui.adsabs.harvard.edu/abs/1986ApJ...304....1P} {304, 1}

\bibitem[\protect\citeauthoryear{{Peterson}}{{Peterson}}{1993}]{Peterson1993}
{Peterson} B.~M.,  1993, \mn@doi [\pasp] {10.1086/133140}, \href
  {https://ui.adsabs.harvard.edu/abs/1993PASP..105..247P} {105, 247}

\bibitem[\protect\citeauthoryear{{Peterson}, {Wanders}, {Horne}, {Collier},
  {Alexander}, {Kaspi}  \& {Maoz}}{{Peterson} et~al.}{1998}]{Peterson1998}
{Peterson} B.~M.,  {Wanders} I.,  {Horne} K.,  {Collier} S.,  {Alexander} T.,
  {Kaspi} S.,   {Maoz} D.,  1998, \mn@doi [\pasp] {10.1086/316177}, \href
  {https://ui.adsabs.harvard.edu/abs/1998PASP..110..660P} {110, 660}

\bibitem[\protect\citeauthoryear{{Peterson} et~al.,}{{Peterson}
  et~al.}{2004}]{Peterson2004}
{Peterson} B.~M.,  et~al., 2004, \mn@doi [\apj] {10.1086/423269}, \href
  {https://ui.adsabs.harvard.edu/abs/2004ApJ...613..682P} {613, 682}

\bibitem[\protect\citeauthoryear{{Planck Collaboration} et~al.,}{{Planck
  Collaboration} et~al.}{2016}]{Planck2016}
{Planck Collaboration} et~al., 2016, \mn@doi [\aap]
  {10.1051/0004-6361/201525830}, \href
  {https://ui.adsabs.harvard.edu/abs/2016A&A...594A..13P} {594, A13}

\bibitem[\protect\citeauthoryear{{Poindexter} \& {Kochanek}}{{Poindexter} \&
  {Kochanek}}{2010}]{Poindexter2010}
{Poindexter} S.,  {Kochanek} C.~S.,  2010, \mn@doi [\apj]
  {10.1088/0004-637X/712/1/658}, \href
  {https://ui.adsabs.harvard.edu/abs/2010ApJ...712..658P} {712, 658}

\bibitem[\protect\citeauthoryear{{Rathna Kumar} et~al.,}{{Rathna Kumar}
  et~al.}{2013}]{RathnaKumar2013}
{Rathna Kumar} S.,  et~al., 2013, \mn@doi [\aap] {10.1051/0004-6361/201322116},
  \href {https://ui.adsabs.harvard.edu/abs/2013A&A...557A..44R} {557, A44}

\bibitem[\protect\citeauthoryear{{Refsdal}}{{Refsdal}}{1964}]{Refsdal1964}
{Refsdal} S.,  1964, \mn@doi [\mnras] {10.1093/mnras/128.4.295}, \href
  {https://ui.adsabs.harvard.edu/abs/1964MNRAS.128..295R} {128, 295}

\bibitem[\protect\citeauthoryear{{Salpeter}}{{Salpeter}}{1964}]{Salpeter1964}
{Salpeter} E.~E.,  1964, \mn@doi [\apj] {10.1086/147973}, \href
  {https://ui.adsabs.harvard.edu/abs/1964ApJ...140..796S} {140, 796}

\bibitem[\protect\citeauthoryear{{Sanders}, {Phinney}, {Neugebauer}, {Soifer}
  \& {Matthews}}{{Sanders} et~al.}{1989}]{Sanders1989}
{Sanders} D.~B.,  {Phinney} E.~S.,  {Neugebauer} G.,  {Soifer} B.~T.,
  {Matthews} K.,  1989, \mn@doi [\apj] {10.1086/168094}, \href
  {https://ui.adsabs.harvard.edu/abs/1989ApJ...347...29S} {347, 29}

\bibitem[\protect\citeauthoryear{{Shapiro}}{{Shapiro}}{1964}]{Shapiro1964}
{Shapiro} I.~I.,  1964, \mn@doi [\prl] {10.1103/PhysRevLett.13.789}, \href
  {https://ui.adsabs.harvard.edu/abs/1964PhRvL..13..789S} {13, 789}

\bibitem[\protect\citeauthoryear{{Shen} et~al.,}{{Shen}
  et~al.}{2015}]{Shen2015}
{Shen} Y.,  et~al., 2015, \mn@doi [\apjs] {10.1088/0067-0049/216/1/4}, \href
  {https://ui.adsabs.harvard.edu/abs/2015ApJS..216....4S} {216, 4}

\bibitem[\protect\citeauthoryear{{Sluse} \& {Tewes}}{{Sluse} \&
  {Tewes}}{2014}]{Sluse2014}
{Sluse} D.,  {Tewes} M.,  2014, \mn@doi [\aap] {10.1051/0004-6361/201424776},
  \href {https://ui.adsabs.harvard.edu/abs/2014A&A...571A..60S} {571, A60}

\bibitem[\protect\citeauthoryear{{Sluse} et~al.,}{{Sluse}
  et~al.}{2003}]{Sluse2003}
{Sluse} D.,  et~al., 2003, \mn@doi [\aap] {10.1051/0004-6361:20030904}, \href
  {https://ui.adsabs.harvard.edu/abs/2003A&A...406L..43S} {406, L43}

\bibitem[\protect\citeauthoryear{{Starkey}, {Horne}  \& {Villforth}}{{Starkey}
  et~al.}{2016a}]{CREAM}
{Starkey} D.~A.,  {Horne} K.,   {Villforth} C.,  2016a, \mn@doi [\mnras]
  {10.1093/mnras/stv2744}, \href
  {https://ui.adsabs.harvard.edu/abs/2016MNRAS.456.1960S} {456, 1960}

\bibitem[\protect\citeauthoryear{{Starkey}, {Horne}  \& {Villforth}}{{Starkey}
  et~al.}{2016b}]{Starkey2016}
{Starkey} D.~A.,  {Horne} K.,   {Villforth} C.,  2016b, \mn@doi [\mnras]
  {10.1093/mnras/stv2744}, \href
  {https://ui.adsabs.harvard.edu/abs/2016MNRAS.456.1960S} {456, 1960}

\bibitem[\protect\citeauthoryear{{Sun}, {Grier}  \& {Peterson}}{{Sun}
  et~al.}{2018}]{PyCCF}
{Sun} M.,  {Grier} C.~J.,   {Peterson} B.~M.,  2018, {PyCCF: Python Cross
  Correlation Function for reverberation mapping studies} (\mn@eprint {ascl}
  {1805.032})

\bibitem[\protect\citeauthoryear{{Tewes} et~al.,}{{Tewes}
  et~al.}{2012}]{Tewes2012}
{Tewes} M.,  et~al., 2012, The Messenger, \href
  {https://ui.adsabs.harvard.edu/abs/2012Msngr.150...49T} {150, 49}

\bibitem[\protect\citeauthoryear{{Tewes}, {Courbin}  \& {Meylan}}{{Tewes}
  et~al.}{2013}]{Tewes2013}
{Tewes} M.,  {Courbin} F.,   {Meylan} G.,  2013, \mn@doi [\aap]
  {10.1051/0004-6361/201220123}, \href
  {https://ui.adsabs.harvard.edu/abs/2013A&A...553A.120T} {553, A120}

\bibitem[\protect\citeauthoryear{{Williams} et~al.,}{{Williams}
  et~al.}{2021}]{Williams2021}
{Williams} P.~R.,  et~al., 2021, \mn@doi [\apj] {10.3847/1538-4357/abe943},
  \href {https://ui.adsabs.harvard.edu/abs/2021ApJ...911...64W} {911, 64}

\bibitem[\protect\citeauthoryear{{Wisotzki}, {Schechter}, {Bradt},
  {Heinm{\"u}ller}  \& {Reimers}}{{Wisotzki} et~al.}{2002}]{Wisotzki2002}
{Wisotzki} L.,  {Schechter} P.~L.,  {Bradt} H.~V.,  {Heinm{\"u}ller} J.,
  {Reimers} D.,  2002, \mn@doi [\aap] {10.1051/0004-6361:20021213}, \href
  {https://ui.adsabs.harvard.edu/abs/2002A&A...395...17W} {395, 17}

\bibitem[\protect\citeauthoryear{{Yu}, {Kochanek}, {Peterson}, {Zu}, {Brandt},
  {Cackett}, {Fausnaugh}  \& {McHardy}}{{Yu} et~al.}{2020}]{Yu2020}
{Yu} Z.,  {Kochanek} C.~S.,  {Peterson} B.~M.,  {Zu} Y.,  {Brandt} W.~N.,
  {Cackett} E.~M.,  {Fausnaugh} M.~M.,   {McHardy} I.~M.,  2020, \mn@doi
  [\mnras] {10.1093/mnras/stz3464}, \href
  {https://ui.adsabs.harvard.edu/abs/2020MNRAS.491.6045Y} {491, 6045}

\bibitem[\protect\citeauthoryear{{Zu}, {Kochanek}  \& {Peterson}}{{Zu}
  et~al.}{2011}]{JAVELIN}
{Zu} Y.,  {Kochanek} C.~S.,   {Peterson} B.~M.,  2011, \mn@doi [\apj]
  {10.1088/0004-637X/735/2/80}, \href
  {https://ui.adsabs.harvard.edu/abs/2011ApJ...735...80Z} {735, 80}

\makeatother
\end{thebibliography}







\bsp	
\label{lastpage}
\end{document}